\documentclass[notitlepage,showpacs,aps,floatfix,prb,reprint,superscriptaddress]{revtex4-1}
\usepackage[colorlinks=true,
            linkcolor=magenta,
            urlcolor=magenta,
            citecolor=magenta]{hyperref}
\usepackage{graphicx}
\usepackage{xfrac}
\usepackage{amssymb}
\usepackage{tikz}
\usepackage{url}
\usepackage{amsfonts}
\usepackage{amsmath}
\usepackage[utf8]{inputenc}
\usepackage{mathtools}
\usepackage{bm}
\usepackage{array}
\usepackage{physics}
\usepackage{amsmath,empheq}
\usepackage{catchfile}
\usepackage{pdfsync}
\usepackage{multirow}
\usepackage{tabularx}
\usepackage[resetlabels]{multibib}
\newcites{sinfo}{References}
\usepackage{layouts}
\usepackage{todonotes}

\usepackage{bm}
\usepackage{amsfonts}
\usepackage{physics}
\usepackage{xfrac}
\usepackage[makeroom]{cancel} 

\newcommand{\bu}{\mathbf{u}}
\newcommand{\bv}{\mathbf{v}}
\newcommand{\br}{\mathbf{r}}
\newcommand{\bq}{\mathbf{q}}

\newcommand{\bt}{\bm{\tau}}

\newcommand{\bk}{\mathbf{k}}

\newcommand{\bR}{\mathbf{R}}

\newcommand{\cS}{\mathcal{S}}
\newcommand{\Sq}{\mathcal{S}\mathbf{q}}
\newcommand{\ka}{\kappa\alpha}
\newcommand{\pk}{p\kappa}

\newcommand{\kRe}{\mathbf{k}\cdot\mathbf{R}_e}
\newcommand{\dVq}{\partial_{\mathbf{q},\kappa\alpha}V(\mathbf{r})}

\newcommand{\dVRp}{\partial_{\mathbf{R}_p,\kappa\alpha}V(\mathbf{r})}
\newcommand{\symop}{\{\mathcal{S}|\mathbf{v}\}}
\newcommand{\isymop}{\{\mathcal{S}|\mathbf{v}\}^{-1}}

\definecolor{mypurple}{RGB}{179,106,226}

\newcommand{\hReRp}{h^{\ka}_{ij}(\mathbf{R}_e,\mathbf{R}_p)}

\def\bkf{\bk_{\text{f}}}
\def\bqf{\bq_{\text{f}}}
\def\bkc{\bk_{\text{c}}}
\def\bqc{\bq_{\text{c}}}

\bibpunct{[}{]}{,}{n}{}{}

\usepackage[final,color,notref,notcite]{showkeys}
\usepackage{seqsplit}
\usepackage{xstring}
\definecolor{labelkey}{rgb}{1.0,0.0,0.0}

\renewcommand*\showkeyslabelformat[1]{ \noexpandarg \StrSubstitute{\(\{\)#1\(\}\)}{ }{\textvisiblespace}[\TEMP] \parbox[t]{\marginparwidth}{\raggedright\normalfont\small\ttfamily\expandafter\seqsplit\expandafter{\TEMP}}}

\begin{document}

\title{\textit{Ab Initio} Electron-Phonon Interactions Using Atomic Orbital Wavefunctions}
\author{Luis A. Agapito} 
\affiliation{Department of Applied Physics and Materials Science, Steele Laboratory, California Institute of Technology, Pasadena, California 91125, United States}
\author{Marco Bernardi} 
\affiliation{Department of Applied Physics and Materials Science, Steele Laboratory, California Institute of Technology, Pasadena, California 91125, United States}
\date{\today}

\begin{abstract}
The interaction between electrons and lattice vibrations determines key physical properties of materials, including their electrical and heat transport, excited electron dynamics, phase transitions, and superconductivity. 
We present a new \textit{ab initio} method that employs atomic orbital (AO) wavefunctions to compute the electron-phonon (e-ph) interactions in materials and interpolate the e-ph coupling matrix elements to fine Brillouin zone grids.
We detail the numerical implementation of such AO-based e-ph calculations, and benchmark them against direct density functional theory calculations and Wannier function (WF) interpolation.  
The key advantages of AOs over WFs for e-ph calculations are outlined. Since AOs are \textit{fixed} basis functions associated with the atoms,  
they circumvent the need to generate a material-specific localized basis set with a trial-and-error approach, as is needed in WFs. 
Therefore, AOs are ideal to compute e-ph interactions in chemically and structurally complex materials for which WFs are challenging to generate, 
and are also promising for high-throughput materials discovery. 
While our results focus on AOs, the formalism we present generalizes e-ph calculations to arbitrary localized basis sets, with WFs recovered as a special case. 
\end{abstract}
\maketitle

%
%
\section{Introduction}
\vspace{-20pt}
Electron-phonon (e-ph) interactions are central to modeling materials properties. While not yet mainstream, \textit{ab initio} e-ph calculations are becoming a key component of computational materials science and condensed matter physics \cite{Bernardi-review, Giustino2017}. 
A key technical challenge of these calculations is obtaining the e-ph coupling matrix elements for different electronic states and phonon modes, within the framework of density functional theory (DFT) and density functional perturbation theory (DFPT)~\cite{Gonze1997_DFPT,Baroni2001DFPT,Savrasov1994eph,Savrasov1996eph, Deinzer}. %
An example are calculations of charge carrier dynamics, which require evaluating the e-ph matrix elements for a large number of electron and phonon wavevectors in the Brillouin zone (BZ). In this and other cases, interpolation of the e-ph matrix elements is essential to achieve numerical convergence \cite{Zhou2016eph_polar, Bernardi-noble,Bernardi2014Si,Bernardi2015GaAs}. 
Previous work~\cite{Giustino2007EPW,Calandra2010Wannier} has shown that interpolation of the e-ph matrix elements can be achieved using maximally localized Wannier functions (WFs) \cite{Marzari2012MLWF}. 
This approach has been successfully employed in recent calculations of e-ph scattering, charge transport, and excited carrier dynamics in semiconductors and metals \cite{Bernardi-noble,Bernardi2014Si,Bernardi2015GaAs,Zhou2016eph_polar,Jhalani2017GaN,Bernardi2015Metals,Li2015eph_converge,Nienen-2018}.\\ 
\indent
However, WF-based e-ph interpolation requires \textit{generating} WFs that can accurately interpolate the bandstructure and e-ph matrix elements. %
While WF generation is straightforward for simple metals and $sp$-bonded semiconductors, it is a trial-and-error approach that becomes challenging for structurally complex systems such as surfaces, interfaces, nanostructures and large supercells, in which the required initial guess for constructing the WFs is not apparent. Similar considerations hold for chemically complex materials with $d$ and $f$ electrons. 
For this and other technical reasons, \textit{ab initio} e-ph calculations have so far focused on relatively simple materials.\\
\indent 
%
%
The fast decay of e-ph interactions in real space is key to WF interpolation of the e-ph matrix elements~\cite{Giustino2007EPW}. 
As a result, any localized basis set can in principle be employed to compute the e-ph matrix elements, 
and, provided they decay rapidly in real space, to interpolate them to arbitrarily fine BZ grids. 
The advantage of WFs is that they constitute a minimal basis set that can accurately interpolate the bandstructure. Localized basis sets such as Gaussian or atomic orbitals (AOs) typically require a number of basis functions in excess of the occupied bands to accurately represent valence and conduction states. 
Yet, a key advantage of these localized basis sets, which are routinely used in quantum chemistry codes, is that they are \textit{fixed}, in the sense that they can be obtained once and stored in a database for future use; this  circumvents the challenge of generating the localized basis set for each new material, as is the case with WFs.\\
\indent   
Recent work has shown that one can use a finite AO basis set to represent the electronic Hamiltonian 
and accurately interpolate an adjustable number of electronic bands~\cite{Agapito_2013_projectionsPRB,Agapito2016TightBinding}.  
Since the accuracy of band structure interpolation obtained with this AO-based method is similar to that of WFs, one may wonder whether AOs $-$ or in fact, any other localized basis set $-$ are also suitable for computing and interpolating the e-ph matrix elements. 
The vision is that using a fixed basis set would automate e-ph interpolation, turning it into a tractable problem that is limited only by computational resources.\\
\indent
Here we present a new method for computing and interpolating the e-ph coupling matrix elements. 
Our approach employs a fixed AO basis set and achieves an accuracy similar to that of WF-based e-ph calculations.  
While the accuracy of our method can be improved systematically by increasing the size of the basis set, we find that a double-$\zeta$ polarized AO basis suffices to accurately reproduce the e-ph matrix elements computed directly with DFT plus DFPT or interpolated with WFs. Our work focuses on AOs, but the formalism we present generalizes e-ph calculations to arbitrary localized basis sets, and we show how WFs can be recovered as a special case. 
Since our approach removes the trial-and-error steps needed to build the localized basis set, interpolation of the e-ph matrix elements $-$ and the related charge carrier dynamics calculations $-$ appear possible for complex materials, surfaces, nanostructures and large supercells, for which WFs are challenging to generate. 
Lastly, since most quantum chemistry methods employ localized basis sets, e-ph calculations based on AOs can be more easily interfaced with accurate post-Hartree-Fock \textit{ab initio} methods \cite{Garnet-1,Garnet-2}, thus opening new possibilities for computing e-ph interactions in strongly correlated materials. Taken together, our work opens new avenues for e-ph calculations in complex materials.  
%
%
%
%
%
%
\section{Methodology}
\vspace{-10pt} %
The e-ph interaction is quantified, within the framework of many-body perturbation theory, by the e-ph matrix elements~\cite{Bernardi-review} 
\begin{equation}\label{eq:g}
g_{mn\nu}(\bk,\bq)\!=\! \left( \frac{\hbar}{2 \omega_{\nu \mathbf{q}}} \right)^{\!\sfrac{1}{2}} 
\!\mel{\psi_{m\bk+\bq}(\br)}
{\Delta_{\nu \mathbf{q}} V(\br)}
{\psi_{n\bk}(\br)}, 
\end{equation}
which represent the transition amplitude from a Bloch electronic state with band index $n$ and crystal momentum $\mathbf{k}$ to a Bloch state with quantum numbers $m$ and $\mathbf{k}+\mathbf{q}$, mediated by the emission or absorption of a phonon with mode index $\nu$ and crystal momentum $\mathbf{q}$. %
All the physical quantities in Eq.~\ref{eq:g} can be computed \textit{ab initio}, the electron wavefunctions $\psi_{n\bk}(\br)$ and $\psi_{m\bk+\bq}(\br)$ using DFT, and the phonon dispersions $\omega_{\nu \mathbf{q}}$ and eigenvectors $e_{\nu\bq}^{\ka}$ (where $\kappa$ labels the atom and $\alpha$ the Cartesian direction) using DFPT~\cite{Baroni2001DFPT}. 
In Eq.~\ref{eq:g}, the perturbation potential induced by a phonon with mode $\nu$ and crystal momentum $\mathbf{q}$ is defined as (see Appendix~\ref{sec:appendixA})~\cite{Bernardi-review}:
\begin{equation}
\label{eq:pertpot}
\Delta_{\nu\bq} V(\br) = \sum_{\ka} \frac{1}{\sqrt{M_{\kappa}}} e_{\nu\bq}^{\ka} \partial_{\bq,\ka}V(\br),
\end{equation} 
where $\partial_{\bq,\ka}V(\br)$ is proportional to the derivatives of the Kohn-Sham potential $V(\br)$ at position $\br$~\footnote{Although the Kohn-Sham potential is non-local when using pseudopotentials, we will use the simplified notation $V(\br)$ to denote it, and warn the reader about the role of the non-local part of the potential when relevant.} 
with respect to changes in the atomic positions $R_{p\kappa\alpha}$ of the atom $\kappa$ (with mass $M_\kappa$) along direction $\alpha$ in the unit cell $p$ located at lattice vector $\mathbf{R}_p$
(in a crystal with periodic boundary conditions and $N_p$ unit cells):
\begin{equation}
\label{eq:partial-pertpot}
\partial_{\bq,\ka}V(\br) = \sum_{\,\,\mathbf{R}_p} e^{i \bq \cdot \mathbf{R}_p} \frac{\partial V(\br)}{\partial R_{p\kappa\alpha}}.
\end {equation}
This perturbation potential is computed using DFPT~\cite{Baroni2001DFPT}. 
\indent
In a basis set of AOs $\phi_j(\br)$, where $j$ is a collective label for the AO quantum numbers, we define the Bloch sums 
\begin{equation}
\label{eq:bsao}
\Phi_{j \bk}(\br) = \frac{1}{\sqrt{N_e}}\sum_{\bR_e} e^{i\kRe} \phi_j(\br-\bR_e).
\end{equation}
where $N_e$ and $\bR_e$ are the number and position of the unit cells in a crystal with periodic boundary conditions. 
%
The DFT electron wavefunctions can be \textit{approximated} with an expansion in Bloch sums: %
\begin{equation}
\psi_{n\bk}(\br) \approx \sum_j A_{jn}^\bk \, \Phi_{j\bk}(\br),
\label{eq:lcao}
\end{equation}
where $A_{jn}^\bk$ are expansion coefficients (in practice, a rectangular matrix $A^\bk$ at each $\bk$-point).
%
%
Using Bloch sums, the e-ph matrix elements in Eq.~\ref{eq:g} can be written as
\begin{equation}\label{eq:g3}
\begin{split}
g_{mn \nu}(\bk,\bq) = \left( \frac{\hbar}{2 \omega_{\nu \bq} }\right)^{\!\sfrac{1}{2}}
\sum_{\ka}\frac{1}{\sqrt{M_{\kappa}}} e_{\nu\bq}^{\ka} \\
\times \sum_{ij} (A^{\bk+\bq}_{im})^*A^{\bk}_{jn}
\,h^{\ka}_{ij}(\bk,\bq),
\end{split}
\end{equation}
where $h^{\ka}_{ij}(\bk,\bq)$ is the matrix element of the e-ph perturbation potential in the AO Bloch sums basis, 
\begin{equation}\label{eq:h-ka-ij}
h^{\ka}_{ij}(\bk,\bq)=
\mel{\Phi_{i\bk+\bq}(\br)}
{\partial_{\bq,\ka} V(\br)}
{\Phi_{j\bk}(\br)}.
\end{equation}
%
%
\begin{figure*}[t]
\centering
\includegraphics{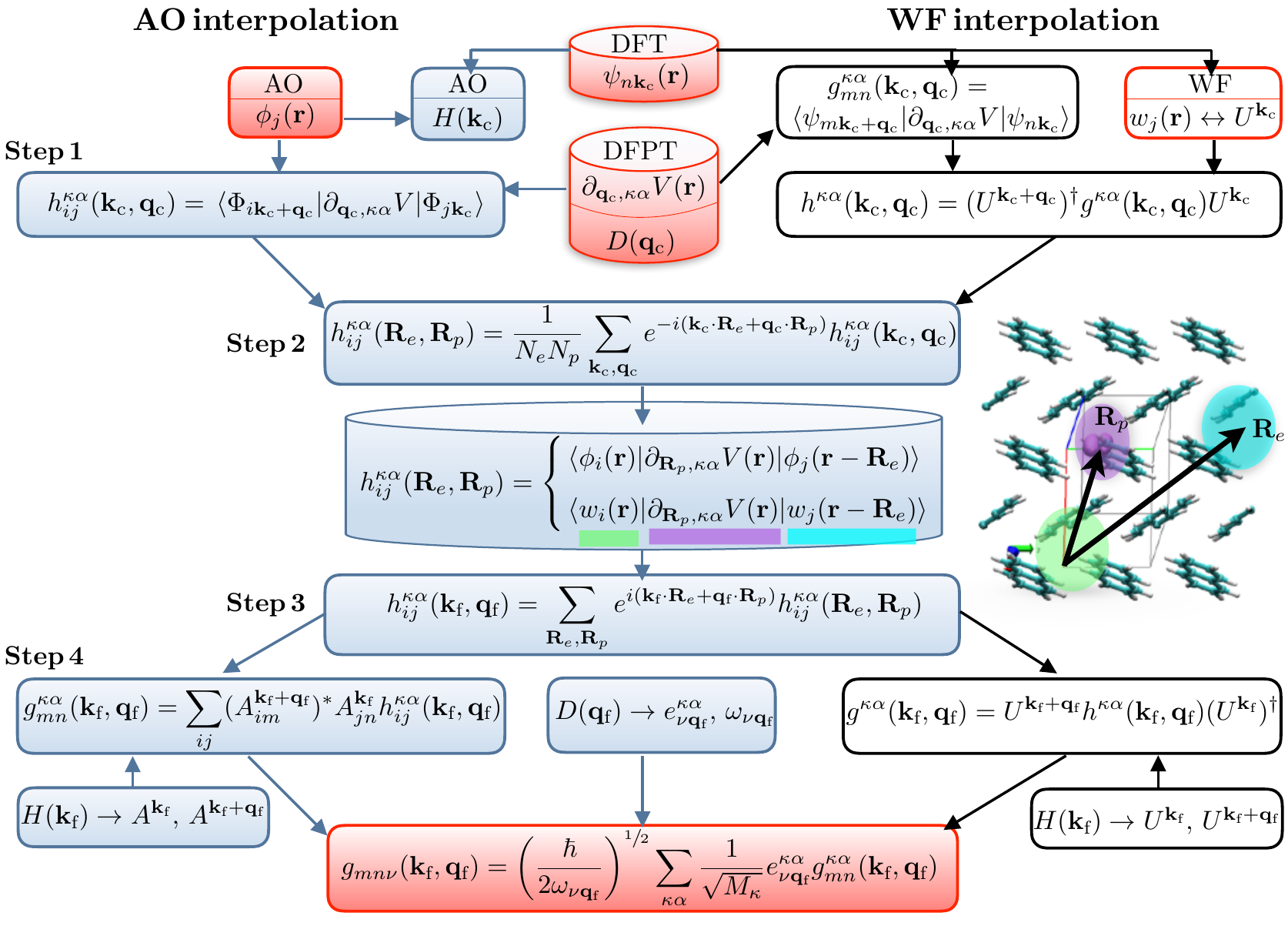}
\caption{Workflow for computing and interpolating the e-ph matrix elements $g_{mn\nu}(\bk,\bq)$ using either AOs (left, blue arrows) or WFs (right, black arrows). The inputs, which are highlighted in red in the top part of the figure, are obtained from DFT (electron wavefunctions and band structure) and DFPT (dynamical matrices and e-ph perturbation potentials). The output, as shown in the red box at the bottom, are the e-ph matrix elements $g_{mn\nu}(\bkf,\bqf)$ at arbitrary fine-grid points $\bkf$ and $\bqf$.  
The inset shows the spatial localization of the e-ph perturbation potential (purple) and localized electronic basis functions (green and cyan), which make the real-space e-ph matrix elements, $h^{\ka}_{ij}(\bR_e,\bR_p)$, decay rapidly, typically over a few unit cells.} %
\label{fig:fig_workflow}
\vspace{-10pt}
\end{figure*}
%
%
\subsection{Interpolation of the e-ph matrix elements}
\vspace{-10pt}
One can show (see Appendix~\ref{sec:appendixB}) that $h^{\ka}_{ij}(\bk,\bq)$ can be written as the double Fourier transform
\begin{equation}
\label{eq:h-doubleft}
h^{\ka}_{ij}(\bk,\bq)= \sum_{\,\,\bR_e,\bR_p} e^{i(\bk\cdot\bR_e + \bq\cdot\bR_p)}
\,h^{\ka}_{ij}(\bR_e,\bR_p)
\end{equation}
of the real-space e-ph perturbation potential in the AO basis,
\begin{equation}
h^{\ka}_{ij}(\bR_e,\bR_p) = \mel{\phi_i(\br)}{\partial_{\bR_p,\ka} V(\br)} {\phi_j(\br-\bR_e)}.
\label{eq:h-Rp-Re}
\end{equation}
where we use the shorthand notation $\partial_{\bR_p,\ka} V(\br)$ for $\partial V (\br)/ \partial R_{p\ka}$ (see Eq.~\ref{eq:partial-pertpot}).
%
%
An important result is that if $h^{\ka}_{ij}(\bR_e,\bR_p)$ decays rapidly in $\bR_e$ and $\bR_p$, 
one can interpolate the e-ph matrix elements on arbitrary fine BZ grids, as explained next.\\
\indent 
Starting from computations of $h^{\ka}_{ij}(\bk,\bq)$ on coarse-grid points $\bk_{\text{c}}$ and $\bq_{\text{c}}$, we compute the inverse double Fourier transform
\begin{equation} \label{eq:h-inverse}
h^{\ka}_{ij}(\bR_e,\bR_p) =
\frac{1}{N_e N_p}\sum_{\bk_{\text{c}},\bq_{\text{c}}} e^{-i( \bk_{\text{c}} \cdot \mathbf{R}_e + \bq_{\text{c}} \cdot \mathbf{R}_p)} 
h^{\ka}_{ij}(\bk_{\text{c}},\bq_{\text{c}}).
\end{equation}
If this quantity decays rapidly in $\bR_e$ and $\bR_p$, then we can interpolate $h^{\ka}_{ij}(\bk,\bq)$ to any pair of fine-grid points $\bk_{\text{f}}$ and $\bq_{\text{f}}$ by carrying out the double Fourier transform
\begin{equation} \label{eq:h-forward}
h^{\ka}_{ij}(\bk_{\text{f}},\bq_{\text{f}}) =
\sum_{\,\,\bR_e,\bR_p} e^{i (\bk_{\text{f}} \cdot \bR_e + \bq_{\text{f}} \cdot \bR_p )}
h^{\ka}_{ij}(\bR_e,\bR_p),
\end{equation}
and from this obtain the e-ph matrix elements $g_{mn\nu}(\bkf,\bqf)$ (using Eq.~\ref{eq:g3}) on arbitrary fine grids.\\
\indent 
%
%
This workflow is detailed in Fig.~\ref{fig:fig_workflow}, which compares the AO-based (this work) and WF-based (Ref.~\cite{Giustino2007EPW}) e-ph interpolation methods.  
%
%
Before the calculation, we collect as input the DFT data (electron wavefunctions and band structure) and DFPT data (dynamical matrices and e-ph perturbation potentials) from calculations, respectively,  
on coarse $\bk$-point and $\bq$-point grids (typically, of size between $4 \times 4 \times 4$ and $12 \times 12 \times 12$ points).
%
%
Note also that the dynamical matrices and e-ph perturbation potentials $\partial_{\bq,\ka}V(\br)$ from DFPT are needed as inputs at all coarse-grid $\bq$-points in the full BZ. However, since DFPT is computationally expensive, we carry out the DFPT calculations only at $\bq$-points in the irreducible BZ wedge, and obtain the dynamical matrices and e-ph perturbation potentials in the full BZ using crystal symmetry operations (see Appendix~\ref{sec:appendixC}). 
%
The last input are the AOs, which can be obtained from databases or, as is done in our work, by solving the radial Schr\"{o}dinger equation for each atomic species.\\ 
\indent
%
%
Before discussing the workflow, let us briefly examine the need to employ fine grids. As discussed above, a typical e-ph calculation employed to compute charge carrier dynamics requires e-ph matrix elements on very fine $\bk$-point and $\bq$-point grids (typically, up to at least $100 \times 100 \times 100$ points) \cite{Zhou2016eph_polar, Bernardi-noble,Bernardi2014Si,Bernardi2015GaAs}. 
Using such dense grids in DFT and DFPT to directly compute $g_{mn\nu}(\bk,\bq)$ is not feasible, both due to the high computational cost of solving the Sternheimer equations of DFPT  
and due to the substantial cost of computing electronic wavefunctions on dense grids with DFT. 
For these reasons, and also because random grids or importance BZ sampling are more convenient in many calculations, interpolation of the e-ph matrix elements is essential.\\
\indent  
%
%
Let us now detail the workflow in Fig.~\ref{fig:fig_workflow}. The first step in the AO calculations consists in forming the AO Bloch sums in Eq.~\ref{eq:bsao} (by projecting the DFT electron wavefunctions~\cite{Agapito_2013_projectionsPRB}), and using Eq.~\ref{eq:h-ka-ij} to compute the e-ph matrix elements $h^{\ka}_{ij}(\bk,\bq)$ in the AO Bloch sum basis for all the coarse-grid $\bk_{\text{c}}$ and $\bq_{\text{c}}$ points.\\ 
\indent
%
In step 2, the matrix elements $h^{\ka}_{ij}(\bR_e,\bR_p)$ are computed using the inverse double Fourier transform in Eq.~\ref{eq:h-inverse}, and stored for later use; this calculation is done for all the lattice vectors $\bR_e$ and $\bR_p$ determined $-$ through the periodic boundary conditions $-$ by the $\bk_{\text{c}}$ and $\bq_{\text{c}}$ coarse grids, respectively.
%
%
The spatial decay of the matrix elements $h^{\ka}_{ij}(\bR_e,\bR_p)$ in both the variables $\bR_e$ and $\bR_p$ needs to be checked in all calculations.  
This decay can be understood from the definition in Eq.~\ref{eq:h-Rp-Re}, which involves three localized functions, $\phi_i(\br)$ centered at the origin, 
$\partial_{\bR_p,\ka} V(\br)$ centered at $\bR_p$, and $\phi_j(\br-\bR_e)$ centered at $\bR_e$. 
Due to the localized nature of the AOs and the e-ph perturbation potential, the integral $h^{\ka}_{ij}(\bR_e,\bR_p)$ decays rapidly as a function of $\bR_e$ and $\bR_p$, as depicted schematically in Fig.~\ref{fig:fig_workflow}. 
%
%
This decay is crucial to reduce the computational cost, since it introduces an upper bound to the number of lattice sites $\bR_e$ and $\bR_p$ at which $\hReRp$ needs to be computed.\\%
\indent
%
%
In step 3 of the AO workflow, we compute $h^{\ka}_{ij}(\bk_{\text{f}},\bq_{\text{f}})$ on an arbitrary fine grid by explicitly carrying out the Fourier transform in Eq.~\ref{eq:h-forward} 
for all pairs of points $\bk_{\text{f}}$ and $\bq_{\text{f}}$ in the fine grid. This procedure is general, and it can be applied to uniform, random, or importance-sampling fine grids. 
Note that one takes advantage of the decay of $h^{\ka}_{ij}(\bR_e,\bR_p)$ beyond a small number of lattice vectors in this step, 
since computing $h^{\ka}_{ij}(\bk_{\text{f}},\bq_{\text{f}})$ at small $\mathbf{k}_{\text{f}}$ and $\bq_{\text{f}}$ vectors would in principle require summing the Fourier transform 
in Eq.~\ref{eq:h-forward} up to correspondingly large lattice vectors $R_e = 2\pi/k_{\text{f}}$ and $R_p = 2\pi/q_{\text{f}}$, respectively, which is not necessary due to the rapid decay.\\
\indent  
%
%
In step 4, we compute the fine-grid e-ph matrix elements in Cartesian coordinates, 
\begin{subequations}\
\label{eq:gkaqfkf}
\begin{align}
g_{mn}^{\ka} (\bk_\text{f},\bq_\text{f}) &= \mel{\psi_{m\bkf+\bqf}(\br)} {\partial_{\bqf,\ka} V(\br)} {\psi_{n\bk_\text{f}}(\br)}
\label{eq:gkaqfkf-1}\\ 
&= \sum_{ij} (A^{\bk_\text{f}+\bq_\text{f}}_{im})^*A^{\bk_\text{f}}_{jn} \, h^{\ka}_{ij}(\bk_{\text{f}},\bq_{\text{f}}). 
\label{eq:gkaqfkf-2}
\end{align}
\end{subequations}
This transformation requires the important auxiliary task of evaluating the expansion coefficients $A^{\bk}_{jn}$ at the fine grid points $\bk_{\text{f}}$ and $\bk_{\text{f}}+\bq_{\text{f}}$.
These coefficients are the components of the AO Hamiltonian matrix eigenvectors~\cite{Agapito2016TightBinding}. 
To obtain them, the AO Hamiltonian matrices $H(\bk)$ are computed for all points $\bkc$ in the coarse grid \cite{Agapito2016TightBinding}, 
and then interpolated to the fine grid points $\bk_\text{f}$ with two consecutive Fourier transforms (see Appendix~\ref{sec:appendixD}):
\begin{subequations}
\label{eq:H-interp}
\begin{align}
H(\bR_e) &= \frac{1}{N_e}\sum_{\bk_{\text{c}}} e^{-i\bk_{\text{c}} \cdot \bR_e} H( {\bk_{\text{c}}} )
\label{eq:H-interp-a}
\\
H({ \bk_{\text{f}}} ) &= \sum_{\bR_e} e^{i\bk_{\text{f}} \cdot \bR_e} H(\bR_e).
\label{eq:H-interp-b}
\end{align}
\end{subequations}
Following this, the Hamiltonians $H( \bkf )$ are diagonalized to obtain the respective eigenvector matrices $A^{\bkf}$.\\ 
\indent
%
%
The final step in the workflow is to transform the Cartesian-coordinates e-ph matrix elements  
to the e-ph matrix elements $g_{mn\nu}(\bk_{\text{f}},\bq_{\text{f}})$ for each given phonon mode $\nu$. 
%
%
The auxiliary task required to this end is computing and diagonalizing the dynamical matrix $D(\bq)$ at the fine-grid points $\bqf$. 
Starting from the dynamical matrices obtained from DFPT at the coarse-grid points $\bqc$, we compute $D(\bq_{\text{f}})$ using standard Fourier interpolation,
\begin{subequations}
\label{eq:D-interp}
\begin{align}
D(\bR_p) &= \frac{1}{N_p}\sum_{\bq_{\text{c}}} e^{-i\bq_{\text{c}} \cdot \bR_p} D(\bq_{\text{c}}) 
\label{eq:D-interp-a}
\\
D(\bq_{\text{f}})  &= \sum_{\bR_p} e^{i\bq_{\text{f}} \cdot \bR_p} D(\bR_p).
\label{eq:D-interp-b}
\end{align}
\end{subequations} 
%
%
After diagonalizing $D(\bq_{\text{f}})$, the phonon frequencies $\omega_{\nu\bqf}$ and eigenvectors $e_{\nu\bqf}^{\ka}$ are employed to obtain 
\begin{equation}
\label{eq:gcart-gnu}
g_{mn\nu} (\bkf,\bqf) = \left( \frac{\hbar}{2 \omega_{\nu \bqf}} \right)^{\!\sfrac{1}{2}} \sum_{\ka} \frac{1}{\sqrt{M_\kappa}} e_{\nu\bqf}^{\ka} \,g_{mn}^{\ka} (\bk_\text{f},\bq_\text{f}).
\end{equation}
This step completes the AO interpolation of the e-ph matrix elements.

%
%
\subsection{Comparison with WF e-ph interpolation}\label{sec:compareWFmethod}
\vspace{-10pt}
The workflow for the WF-based e-ph interpolation is also shown in Fig.~\ref{fig:fig_workflow}, and discussed briefly to compare with the AO method.
The WF scheme can be considered as a particular case of the AO interpolation described above.  
The WFs centered at $\bR_e$ are defined as 
\begin{equation}\label{eq:wf}
w_j(\br-\bR_e) = \frac{1}{\sqrt{N_e}} \sum_{n\bk}e^{-i\kRe}U^{\bk}_{nj}\psi_{n\bk}(\br)
\end{equation}
and determined by finding the unitary matrices $U^{\bk}$ that maximize the WF spatial localization \cite{Marzari2012MLWF}. 
To make a parallel with AOs, we introduce Bloch sums of WFs, 
\begin{equation}
\label{wf-bsum}
W_{j\bk}(\br) = \frac{1}{\sqrt{N_e}}\sum_{\,\bR_e} e^{i\kRe}w_j(\br-\bR_e),
\end{equation} 
which are fully analogous to the AO Bloch sums in Eq.~\ref{eq:bsao}.
%
%
We can thus write the electron wavefunctions as 
\begin{align}
\label{eq:lcwf}
\psi_{n\bk}(\br) &= \frac{1}{\sqrt{N_e}} \sum_{j\bR_e}e^{i\kRe}(U^{\bk}_{nj})^* w_j(\br-\bR_e) \\
                        &= \sum_j \left(U^{\bk}_{nj}\right)^{*} W_{j\bk} (\br),
\end{align}
which highlights the parallel between the WF and AO formalisms since the Wannier matrix elements $\left(U^{\bk}_{nj}\right)^*$ are analogous 
to the AO expansion coefficient $A^{\bk}_{jn}$ in Eq.~\ref{eq:lcao}.\\
\indent
The WF interpolation workflow is almost identical to that for AOs, with an important difference in the first step. 
%
%
Different from the AOs, the DFT electron wavefunctions can be expanded \textit{exactly} in the WF basis set (through Eq.~\ref{eq:lcwf}). %
Therefore, in the first step of the WF interpolation we compute directly $g^{\ka}_{mn}(\bkc,\bqc)= \mel{\psi_{m\bkc+\bqc}(\br)} {\dVq} {\psi_{n\bkc}(\br)}$ on the coarse grids, 
and then obtain $h_{ij}^{\ka} (\bkc,\bqc)$ as (see Appendix~\ref{sec:mlwf})
\begin{equation}
\label{eq:wfhij}
h_{ij}^{\ka} (\bkc,\bqc) = \sum_{mn} (U_{mi}^{\bkc + \bqc})^* g^{\ka}_{mn}(\bkc,\bqc) U_{nj}^{\bkc}
\end{equation}
using the Wannier matrices $U^{\bkc}$. 
%
%
In the WF approach, to consistently fix the phase of the electron wavefunctions and e-ph matrix elements (or their gauge in the case of degenerate electronic states), the WFs and Wannier matrices need to be generated with the same DFT electron wavefunctions employed to compute $g^{\ka}_{mn}(\bkc,\bqc)$. %
%
%
Note how in the AO method computing $g^{\ka}_{mn}(\bkc,\bqc)$ as a first step and then obtaining $h_{ij}^{\ka} (\bkc,\bqc)$ from it would be incorrect $-$ since the expansion of the DFT electron wavefunctions in the AO basis set is only approximate, the phase information of the electron wavefunctions would be lost. When using AO Bloch sums, it is natural to compute $h_{ij}^{\ka} (\bkc,\bqc)$ directly, %
and there is no ambiguity in the phase (or gauge) of $h_{ij}^{\ka} (\bkc,\bqc)$, which is fixed by the definition of the Bloch sums in Eq.~\ref{eq:bsao}.\\%
\indent
%
%
Beyond the first step, the WF e-ph interpolation workflow in Fig.~\ref{fig:fig_workflow} is equivalent to that for AOs. 
Considerations analogous to those discussed above also hold for the WF interpolation of the Hamiltonian, dynamical matrices, and Wannier matrices~\cite{Giustino2007EPW}. 
The AO workflow presented here is general and can be adapted to any localized basis set.%
%
%
%
%
%
%
\section{Results} 
\vspace{-10pt}
%
%
The e-ph calculations using AOs and WFs have been implemented in our code \textsc{Perturbo} \footnote{\url{http://perturbo.caltech.edu/}} 
following the workflows in Fig.~\ref{fig:fig_workflow}. We employ the code for benchmark calculations on silicon and diamond, which are discussed below.
In these calculations, the unitary matrices $U^\bk$ for the WF interpolation are computed with \textsc{Wannier90} \cite{Mostofi2014W90}, 
and the Hamiltonians in the AO basis with \textsc{PyTB} \cite{[][{. \url{http://esl.cecam.org/PyTB}}]Agapito_2013_projectionsPRB}. 
The AO basis sets for Si and C are obtained by solving their atomic radial Schr\"{o}dinger equation with the \texttt{ld1.x} utility of \textsc{Quantum Espresso}~\cite{Giannozzi2009}. 
A double-$\zeta$ polarized basis set is employed, which includes the $n\,s$ and $n\,p$ occupied AOs ($n$=2,$\,$3 for C and Si, respectively), doubling orbitals obtained following Ref.~\cite{Sanchez-Portal1996}, and 3$d$ polarization orbitals, 
for a total of 13 AO basis functions per atom, all of which are pseudized with the norm-conserving procedure. 
The basis set includes unbound orbitals with oscillatory character, which we terminate with an exponential tail to retain the localized character. 
All DFT and DFPT calculations are performed using \textsc{Quantum Espresso}~\cite{Giannozzi2009}. The local density approximation~\cite{Perdew1981LDA} is employed for silicon, 
and the Perdew-Burke-Ernzerhof generalized gradient approximation~\cite{Perdew1996} for diamond. 
Norm-conserving pseudopotentials
\cite{Troullier1991NCPP,Kleinman1982SeparableNCPP}
 from the PSlibrary \cite{[][{. \url{http://www.qe-forge.org/gf/project/pslibrary/}}]DalCorso2014Pseudos}
are used, together with a 60 Ry plane-wave kinetic energy cutoff. 
%
%
%
%
\subsection{Interpolation of the e-ph matrix elements}
\begin{figure}[h] 
\centering 
\includegraphics[width=7.0cm,height=11cm]{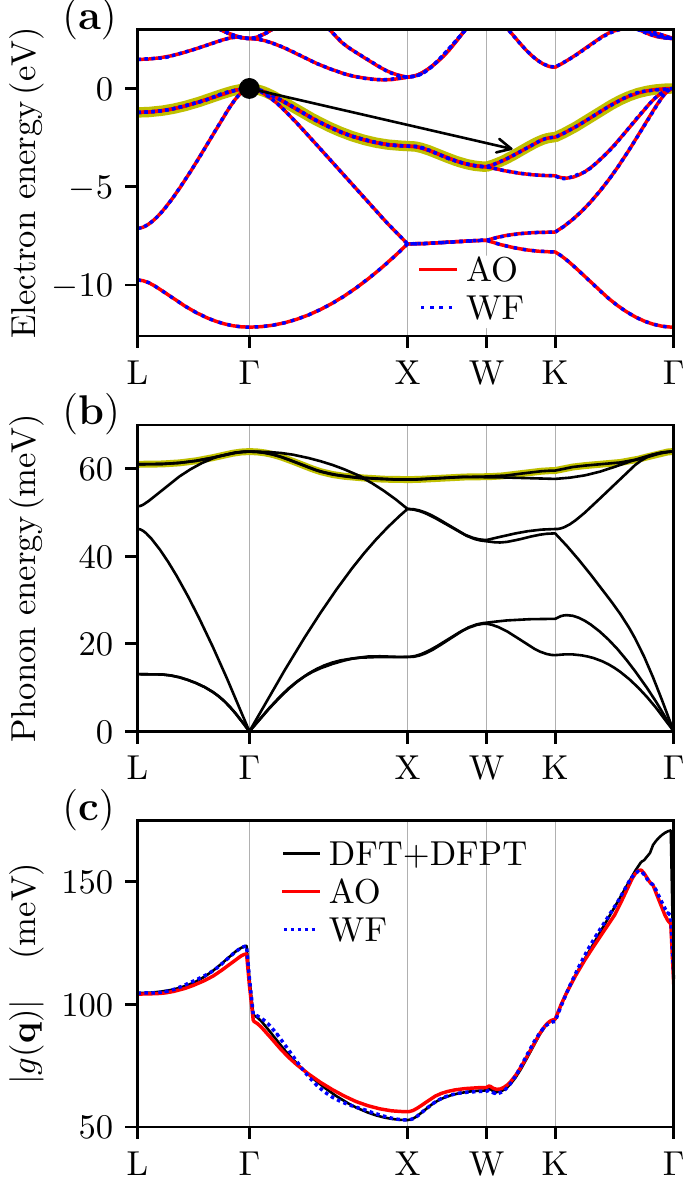}
\caption{Comparison of AO and WF interpolations for silicon. Shown are the interpolated (a) band structure, (b) phonon dispersions, and (c) e-ph matrix elements. 
The interpolated e-ph matrix elements are compared with those computed directly on the fine grid with DFT plus DFPT.  %
The highest valence band highlighted in (a) and the optical phonon mode highlighted in (b) are employed in the e-ph matrix elements calculations in (c). 
As shown schematically in (a), the initial electronic state is fixed at the valence band maximum.} %
\label{fig:fig_gcompare}
\end{figure}

\vspace{-10pt}
%
%
Figure~\ref{fig:fig_gcompare} shows the electronic band structure, phonon dispersions, and e-ph matrix elements in silicon, all interpolated with the AO method. The results are compared with those obtained with WF interpolation (using the same coarse and fine grids) as well as with direct DFT and DFPT calculations. Such direct DFT plus DFPT calculations, in which the wavefunctions and e-ph perturbation potentials entering $g_{mn\nu}(\bk,\bq)$ (see Eq.~\ref{eq:g}) are computed directly on the fine grid with DFT and DFPT, are used as benchmark for the e-ph matrix elements interpolation. To make the comparison quantitative, we compute root-mean-square (RMS) deviations between the different data sets.\\
\indent
For both the AO and WF methods, the interpolated electronic eigenvalues are within $\sim$10 meV of the DFT result throughout the BZ (see Fig. S1 in the Supplemental Material). 
The accuracy of the AO and WF interpolated band structures is comparable, both for the valence and conduction bands, and the band structure interpolation does not pose particular challenges to the AO method. 
The AO interpolation of the DFT electron wavefunctions is more subtle since the AOs are not a complete basis set; this aspect, which is the main challenge in the AO method, is discussed in detail below. 
The phonon dispersions are also accurate both in the AO and WF methods. The interpolation of the dynamical matrices is independent of the chosen localized basis set, so that any small error in the phonon frequencies and eigenvectors is identical in the AO and WF interpolations.\\ %
\indent 
Figure~\ref{fig:fig_gcompare}(c) shows the e-ph matrix elements interpolated on a fine grid using AOs, and compares them to results from WF interpolation and to direct DFT plus DFPT calculations, which are used as benchmark. The interpolated e-ph matrix elements are based on calculations with $12\times12\times12$ $\bk$-point and $6\times6\times6$ $\bq$-point coarse grids. The quantity plotted for the comparison in Fig.~\ref{fig:fig_gcompare}(c) is $g_{mn\nu}(\bkf,\bqf)$ for $\bkf \!=\! 0$ and as a function of $\bqf$ along a high-symmetry path; the initial band $n$ and the final band $m$ are the top valence band, and $\nu$ is fixed to the phonon mode shown in Fig.~\ref{fig:fig_gcompare}(b). 
We find that both the AO and the WF interpolation methods can accurately reproduce the e-ph matrix elements from direct DFT plus DFPT calculations. The discrepancy between the interpolated and directly computed results near $\Gamma$ along the K$-$$\Gamma$ direction is a numerical artifact present in both the AO and WF methods, which is discussed in Section~\ref{sec:jump}. 
We also find that the AO and WF e-ph matrix elements are in excellent agreement with each other, with a small RMS deviation between the two data sets of $\sim$2.1 meV on the chosen high-symmetry path. 
While this difference between the AO and WF interpolated e-ph matrix elements is small and can be safely dismissed, it is important to understand its origin.\\%
\indent
%
%
%
%
%
To this end, we analyze the difference between the directly computed and AO or WF interpolated e-ph matrix elements 
$g_{mn\nu}(\bkf\!=\!0,\bqf)$ for $\bqf$ at several high-symmetry points (see Table~\ref{table:Si-AOvsWF}). 
Results are given for interpolations using coarse grids ($\bqc$,$\bkc$) with sizes $(4^3,12^3)$, $(6^3,12^3)$ and $(8^3,8^3)$ (here and below, $N^3$ is shorthand to indicate $N \!\times N \times\! N$ grids). 
In this particular analysis, we use the same phonon frequencies and eigenvectors for the interpolated and benchmark DFT plus DFPT results, so that errors in the interpolated e-ph matrix elements 
can only be due to the interpolated electron wavefunctions and e-ph perturbation potential (see Eq.~\ref{eq:g}).\\
\indent    
Critical to the accuracy of the interpolation is whether the fine-grid $\bkf$ and $\bqf$ points are also present in the coarse grids. 
For points $\bqf$ that are also present in the DFPT coarse grid, the Fourier interpolation of the e-ph perturbation potential gives exactly the DFPT result. 
%
There is an important difference in how the electronic wavefunctions are interpolated in the AO and WF methods. For points $\bkf$ that are also present in the DFT coarse grid, the WF interpolated wavefunctions are \textit{exactly} equal to the DFT result. By contrast, for AOs the interpolated wavefunction is approximate at all $\bkf$ points, regardless of whether they are present in the coarse grid. Since the AO basis set is incomplete, small errors in reproducing the DFT wavefunctions $-$ especially in the valence regions between the atoms $-$ are expected to result in small interpolation errors in the e-ph matrix elements.\\
\indent   
For the L and X high-symmetry points in Table~\ref{table:Si-AOvsWF}, the fine-grid point $\bqf$ is present in all the three coarse $\bqc$ grids considered, 
so that the only possible source of error at these points is the interpolated electron wavefunction. 
Since the L and X points are also part of all the coarse $\bkc$ grids considered, WF interpolation can reproduce exactly all the quantities entering the e-ph matrix elements. 
Accordingly, there is no discrepancy between the interpolated and direct DFT plus DFPT results for WFs at L and X (see Table~\ref{table:Si-AOvsWF}). 
For AO interpolation, we find a small error of respectively -0.43 and 3.41 meV in the interpolated e-ph matrix elements at L and X, which derives exclusively from the interpolated electron wavefunctions. While the interpolated AO Hamiltonians give accurate eigenvalues at L and X, the accuracy of the wavefunctions is affected by the AO basis set truncation error.\\
\indent
%
The high-symmetry point K in Table~\ref{table:Si-AOvsWF} is present in the coarse grid only for coarse $\bqc$ and $\bkc$ grids with $8^3$ points, for which the WF error vanishes, and the AO error is 0.36 meV. 
However, for coarse grids ($\bqc$, $\bkc$) with size ($4^3$, $12^3$) and ($6^3$, $12^3$), the point K is not present in the coarse $\bkc$ grids employed in DFT, so that errors due exclusively to the interpolated wavefunctions are expected for both WFs and AOs. Accordingly, we find an error of 1.3 meV for AOs and 1.71 meV for WFs when we use the ($4^3$, $12^3$) coarse grid, and 0.19 meV for AOs and -0.69 meV for WFs with the ($6^3$, $12^3$) coarse grid.\\
\indent
This analysis highlights that, when errors are present in both methods, the interpolation error is comparable for the AO and WF approaches. 
This is the case for all fine-grid points that are not present in the coarse grids, and thus for the vast majority of points in a typical calculation. 
These considerations also apply to diamond~(see Table~\ref{table:C-AOvsWF} in the Supplemental Material), for which the errors show the same trends as in silicon.\\
\indent 

%
%
\begin{table}[!hb]
\vspace{-10pt}
\caption{Difference (in meV units) between the interpolated e-ph matrix elements in Fig.~\ref{fig:fig_gcompare}(c) and those computed directly from DFT plus DFPT at three high-symmetry $\bqf$ points. 
The data are for silicon, and both the AO and WF interpolated results are given for several coarse grids.} 
\label{table:Si-AOvsWF} 
\begin{ruledtabular}
\begin{tabular}{c|c|c|c|c}
$\bqf$ point & Method &\multicolumn{3}{c}{Coarse grid size ($\bqc$ grid, $\bkc$ grid)}\\
        &        & ($4^3$, $12^3$) & ($6^3$, $12^3$) & ($8^3$, $8^3$)\\
\hline
\multirow{2}{*}{K = [-$\frac{3}{8}$, $\frac{3}{8}$, 0]} &AO& 1.30 & 0.19  & 0.36\\ 
&WF& 1.71 & -0.69 & 0.00\\
\hline
\multirow{2}{*}{$\!\!\!\!\!$L = [0, $\frac{1}{2}$,0]} &AO& -0.43 & -0.43 &-0.43\\ 
&WF& 0.00 & 0.00 & 0.00  \\
\hline
\multirow{2}{*}{X = [0, $\frac{1}{2}$, $\frac{1}{2}$]} &AO& 3.41 & 3.41  & 3.41\\ 
&WF& 0.00 & 0.00 & 0.00\\
\end{tabular}
\end{ruledtabular}
\vspace{50pt}
\end{table}
%
%
\begin{figure*}[!th]
\centering
\includegraphics[width=17.0cm]{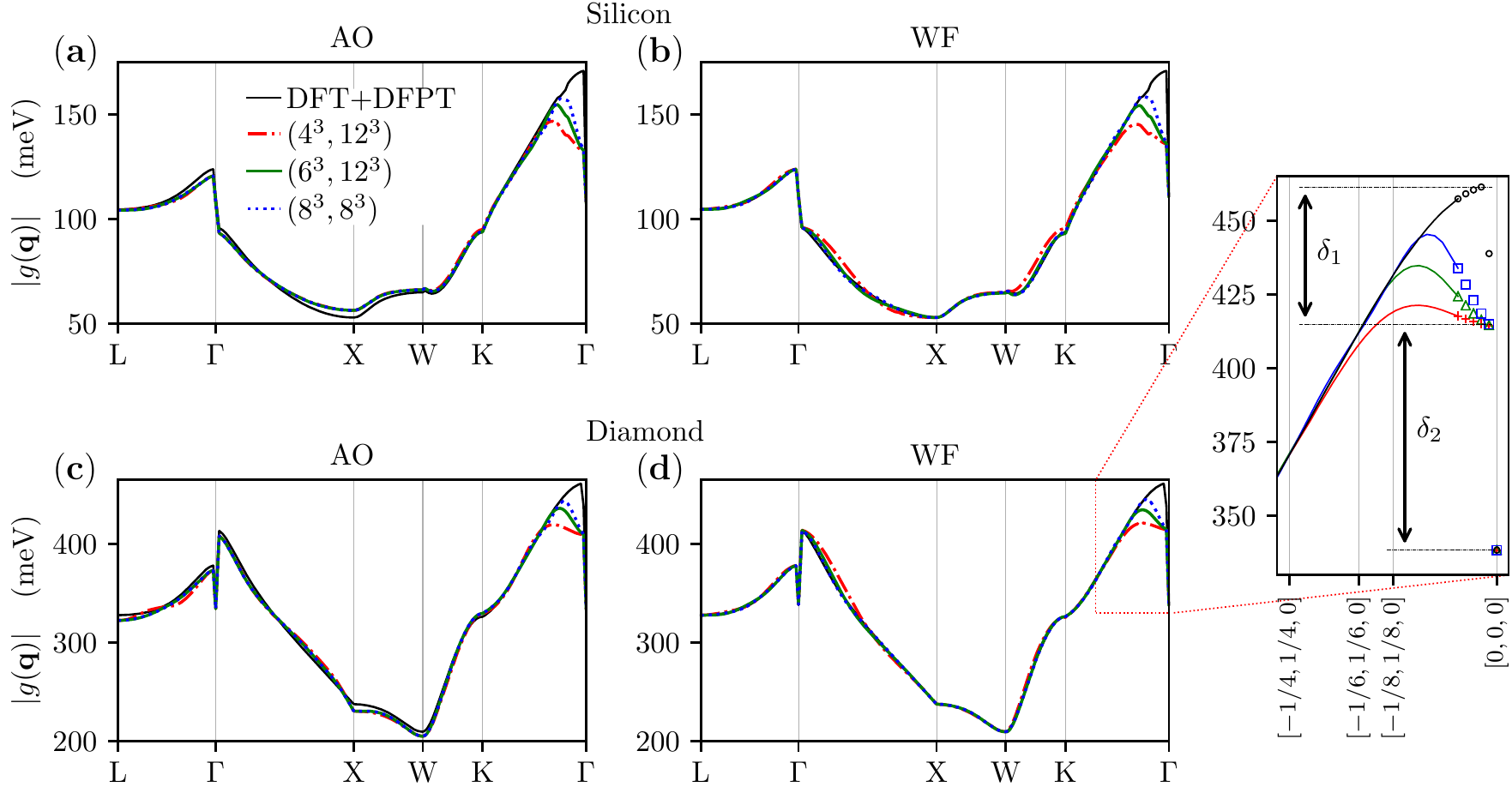}
\caption{Interpolated e-ph matrix elements for (a)-(b) silicon and (c)-(d) diamond, for different coarse grids ($\bqc$, $\bkc$) of size given in the legend. 
For each material, we show in separate panels AO and WF interpolated e-ph matrix elements $g_{mn\nu}(\bkf=0,\bqf)$, for $\bqf$ along a high-symmetry path; the bands and phonon modes are chosen as in Fig.~\ref{fig:fig_gcompare}. The inset zooms into the discrepancy near $\Gamma$ discussed in Section~\ref{sec:jump}.}
\label{fig:fig_gconverge}
\vspace{-15pt}
\end{figure*}
%
Lastly, we analyze the convergence with respect to the coarse grid size, for both silicon and diamond, in Fig.~\ref{fig:fig_gconverge} and Table~\ref{table:convergence}. 
The slow convergence near $\Gamma$ along the K$-$$\Gamma$ direction, which is discussed in Section~\ref{sec:jump}, is not included in this analysis. 
%
We find that the interpolated e-ph matrix elements are converged for coarse $\bqc$ and $\bkc$ grids, respectively, of size $6^3$ and $8^3$ points. 
Using denser coarse grids does not appreciably reduce the interpolation errors (see Table~\ref{table:convergence}). 
For coarse $\bkc$ grids denser than $8^3$ points, the AO interpolated results are nearly unchanged. 
Interestingly, the WF results change as a function of coarse $\bkc$ grid even at convergence, since different coarse $\bkc$ grids correspond to different numbers of exact electron wavefunctions employed in the interpolation. 
For a converged coarse grid ($\bqc$, $\bkc$) of size ($8^3$, $8^3$), %
the AO interpolated e-ph matrix elements exhibit a RMS deviation (compared to direct DFT plus DFPT calculations) of 1.8 meV for silicon and 3.8 meV for diamond, 
versus a smaller RMS deviation of 0.5 meV for silicon and 0.6 meV for diamond for WF interpolation (see Table~\ref{table:convergence}). 
We remark that these RMS deviations, for both WF and AO interpolations, are very small, roughly 1\% of the e-ph matrix elements absolute value. We attribute the slightly lower accuracy of the AO interpolation method 
to the fact that the interpolated wavefunctions are approximate at all grid points when using AOs.\\
\indent

%
%
%
\begin{table}[!ht]
\vspace{-10pt}
\caption{RMS deviations (in meV units) between the interpolated e-ph matrix element in Fig.~\ref{fig:fig_gconverge} and the direct DFT plus DFPT results.  
The error near $\Gamma$ along the K$-$$\Gamma$ path is not included in the RMS deviations.} %
\label{table:convergence} 
\begin{ruledtabular}
\begin{tabular}{c|c|c|c|c}
Material & Method &\multicolumn{3}{c}{Coarse grid size ($\bqc$ grid, $\bkc$ grid)}\\
         &        & ($4^3$, $12^3$) & ($6^3$, $12^3$) & ($8^3$, $8^3$) \\%
\hline
\multirow{2}{*}{Silicon} &AO& 2.0 & 1.8 & 1.8 \\
                         &WF& 2.3 & 0.8 & 0.5 \\
\hline
\multirow{2}{*}{Diamond}  &AO& 4.7 & 3.8 & 3.8 \\
                         &WF& 3.6 & 1.3 & 0.6 \\
\end{tabular}
\end{ruledtabular}
\vspace{-20pt}
\end{table}
%
%
%
\subsection{Spatial decay of the e-ph matrix elements}
\vspace{-10pt}
The AO and WF interpolation workflows in Fig.~\ref{fig:fig_workflow} introduce the e-ph matrix elements in real space and Cartesian coordinates, $\hReRp\!$ (see Eq.~\ref{eq:h-Rp-Re}). 
Their spatial decay is critical to the success of the interpolation procedure, as discussed above. 
%
%
\begin{figure}[!h]
\centering
\includegraphics[width=8.6cm]{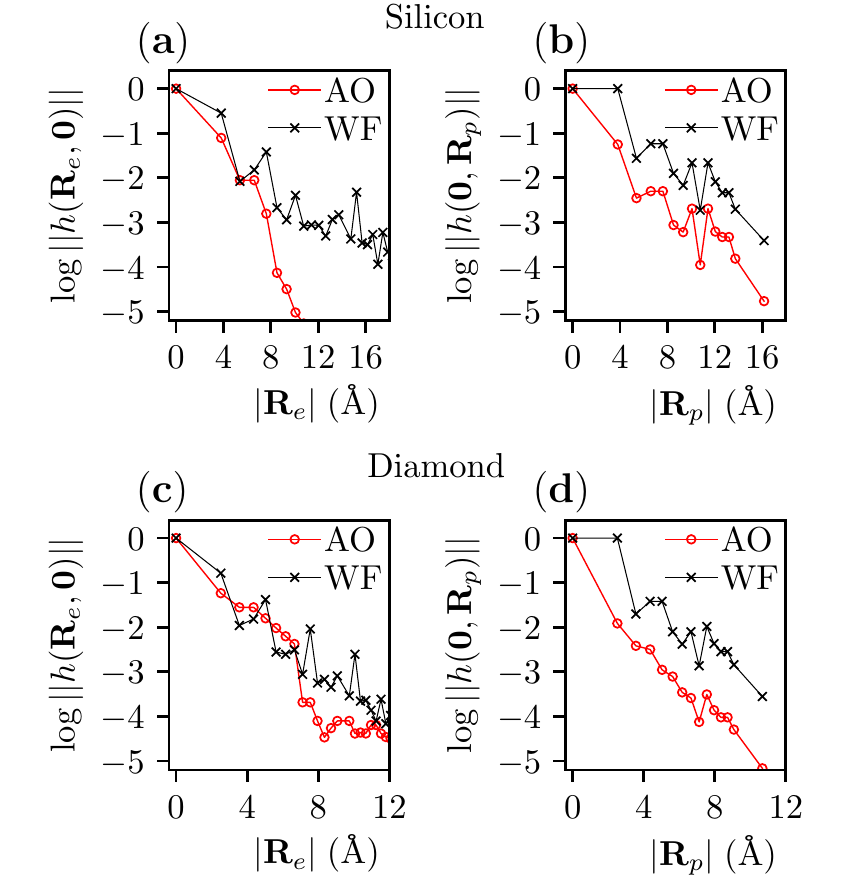}
\caption{Spatial decay of the real-space e-ph matrix elements. The maximum values of the matrix elements, $||h(\bR_e,\bR_p)||$, are normalized to 1 and plotted on a logarithmic scale.  
Panels (a) and (c) show $||h(\bR_e,\bR_p)||$ as a function of $\bR_e$ for $\bR_p = 0$, and panels (b) and (d) as a function of $\bR_p$ for $\bR_e = 0$.  
Results are shown, for both the AO and WF basis sets, for silicon and diamond. The nearly linear trends seen in all plots indicate an approximately exponential decay of the matrix elements over a 2$-$3 unit cell distance of roughly 10 \AA.}
\label{fig:hRpRe_decay}
\end{figure}
To analyze the spatial behavior of $\hReRp\!$, following Ref.~\cite{Giustino2007EPW} we define, for each pair of $\bR_e$ and $\bR_p$ lattice vectors, the matrix element of maximum absolute value as $||h(\bR_e,\bR_p)|| = \max_{\ka,ij}|\hReRp|$. 
Figure~\ref{fig:hRpRe_decay} shows the spatial behavior of $||\hReRp||$ for silicon and diamond, both as a function of $\bR_e$ while keeping $\bR_p = 0$ and as a function of $\bR_p$ while keeping $\bR_e = 0$. 
We find an exponential decay over a few unit cells of these real-space e-ph matrix elements, for both AOs and WFs. This result, which is a consequence of the spatial localization of the WF and AO basis sets, 
establishes that both AOs and WFs are suitable for interpolating the e-ph matrix elements. 
\vspace{20pt}
%
%
%
\subsection{Computation of the e-ph self-energy}
%
%
\vspace{-10pt}
\begin{figure}[h]
\centering
\includegraphics[width=8.6cm,height=12cm]{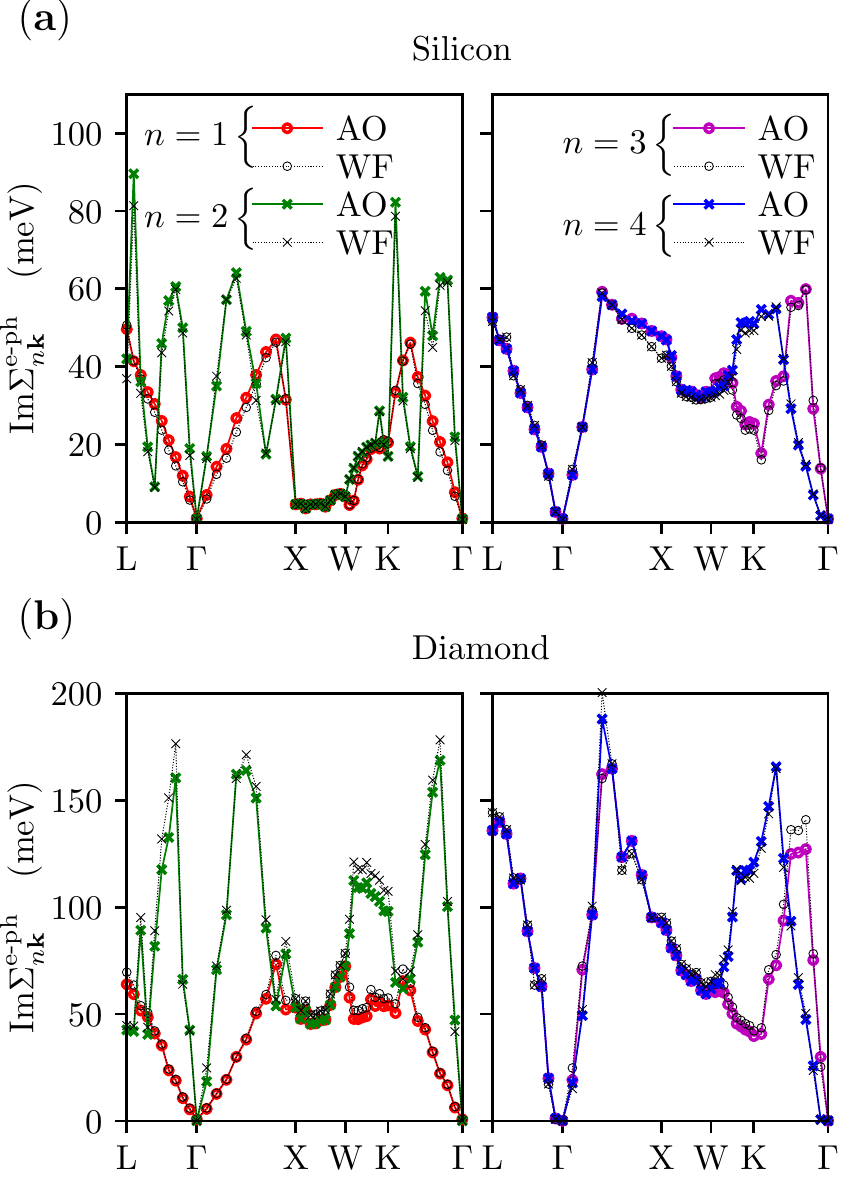}
\caption{Imaginary part of the e-ph self-energy, for (a) silicon and (b) diamond. For each material, we plot $\textrm{Im}\Sigma_{n\bk}$ for the top four valence bands, labeled $n\!=\!1$$-$4 in order of increasing energy, along the shown $\bk$-point path.  
For each band, the AO interpolation results (color-coded curves) are compared with the WF interpolation results (black dashed curves).} 
\label{fig:fig_linewidths}
\end{figure}
%
%
The e-ph scattering rates $\Gamma^{\textrm{e-ph}}_{n\textbf{k}}$ are central quantities in the calculation of charge transport and excited electron dynamics~\cite{Bernardi-noble,Bernardi2014Si,Bernardi2015GaAs,Zhou2016eph_polar,Jhalani2017GaN,Bernardi2015Metals,Li2015eph_converge,Nienen-2018}. In the lowest order of perturbation theory \cite{Bernardi-review}, they read:
\begin{align}
\label{eq:e_scat_rate}
\Gamma^{\textrm{e-ph}}_{n\textbf{k}}=\;\frac{2\pi}{\hbar} & \sum_{m\nu\textbf{q}}|g_{mn\nu}(\textbf{k},\textbf{q})|^2\\[2pt]
\times\big[ & (N_{\nu\textbf{q}}+1-f_{m\textbf{k}+\textbf{q}})\delta(\varepsilon_{n\textbf{k}}-\varepsilon_{m\textbf{k}+\textbf{q}}-\hbar\omega_{\nu\textbf{q}})\nonumber\nonumber\\[5pt]
&+(N_{\nu\textbf{q}}+f_{m\textbf{k}+\textbf{q}})\delta(\varepsilon_{n\textbf{k}}-\varepsilon_{m\textbf{k}+\textbf{q}}+\hbar\omega_{\nu\textbf{q}})\big],\nonumber
\end{align}
where $f_{n\textbf{k}}$ and $N_{\nu\textbf{q}}$ are the electron and phonon occupations, respectively, and the other quantities have been defined above. 
Computing the e-ph scattering rates is a rather stringent test for the AO method because the calculations employ a large number of interpolated e-ph matrix elements (roughly $10^6$ for each $\bk$-point at which $\Gamma_{n\bk}^{\textrm{e-ph}}$ is computed) distributed throughout the BZ. %
The accuracy of the e-ph scattering rates allows us to establish whether the small errors in the AO interpolation of the e-ph matrix elements build up into large discrepancies. 
We compute the e-ph scattering rates with a uniform $90^3$ fine $\bq$-point grid, which is necessary to converge the sum in Eq.~\ref{eq:e_scat_rate}~\cite{Li2015eph_converge,Bernardi2014Si}, 
starting from coarse $6^3$ $\bq$-point and $12^3$ $\bk$-point grids~\cite{Bernardi2014Si}.\\
\indent
%
%
Fig.~\ref{fig:fig_linewidths} shows the e-ph scattering rates in silicon and diamond, expressed as the imaginary part of the e-ph self-energy, $\mathrm{Im} \Sigma^{\textrm{e-ph}}_{n\bk} = (\hbar/2) \Gamma^{\textrm{e-ph}}_{n\textbf{k}}$ (in meV units), for electronic states in the top four valence bands and for $\bk$-points along a BZ high-symmetry path. 
We find that the AO and WF interpolation methods give e-ph scattering rates in very good agreement with each other, 
and that both methods can reproduce the sharp changes of $\mathrm{Im} \Sigma^{\textrm{e-ph}}_{n\bk}$ along the BZ path. 
The RMS deviations between the AO and WF data sets are, for bands $n=1$$-$4 respectively, 1.4, 2.0, 1.8, and 1.8 meV in silicon and 3.4, 6.9, 4.6, and 3.8 meV in diamond. 
These deviations are very small, roughly 1\% of the $\mathrm{Im} \Sigma^{\textrm{e-ph}}_{n\bk}$ values.
The result for band $n\!=\!4$ in diamond along the L$-$$\Gamma$$-$X path agrees with previous calculations 
\footnote{
See Fig.~9d in Ref.~\onlinecite{Giustino2007EPW}, which, as later clarified in Ref.~\onlinecite{Ponce2016EPW}, corresponds to the result for diamond.
}.
We analyze briefly the origin of the small deviations between the two methods. We can rule out the role of the electronic energies since recomputing $\mathrm{Im} \Sigma^{\textrm{e-ph}}_{n\bk}$ with the WF method but with AO interpolated electronic energies (or viceversa, AO computations with WF interpolated electronic energies) leads to negligible changes in the results. This is consistent with the excellent match between the AO- and WF-interpolated band structures.
We also rule out the phonon frequencies and e-ph perturbation potentials, which are the same in both methods.  
We thus conclude that, similar to what we find above for the e-ph matrix elements, the primary source of discrepancy between the AO and WF e-ph self-energy 
is the interpolation of the electron wavefunctions, which is affected by the AO basis set truncation error. 
%
%
%
%
\subsection{Interpolation in the $q \to 0$ limit}
\label{sec:jump}
\vspace{-10pt}
The slow convergence of the e-ph interpolation near $\Gamma$ along the K$-$$\Gamma$ direction in Fig.~\ref{fig:fig_gconverge} is a consequence of the treatment of $q \to 0$ (long-wavelength) perturbations in DFPT~\cite{Baroni2001DFPT}. This point, which has been discussed in Ref.~\cite{Calandra2010Wannier}, is briefly outlined here. 
We define the lattice-periodic part of the e-ph perturbation potential in Eq.~\ref{eq:partial-pertpot}:
\begin{align}
\partial_{\bq,\ka}v(\br) &= e^{-i\bq \cdot \br}\, \partial_{\bq,\ka}V(\br) \\
&= \sum_{\,\,\mathbf{R}_p} e^{-i \bq \cdot (\br - \mathbf{R}_p)} \frac{\partial V(\br)}{\partial R_{p\kappa\alpha}}.
\label{periodic-eph-pert}
\end{align}
This lattice-periodic perturbation potential, which is the quantity stored to disk in the DFPT implementation of \textsc{Quantum Espresso}, 
is the sum of a Coulomb (i.e., electrostatic) and an exchange-correlation contribution,
\begin{equation}
\partial_{\bq,\ka}v(\br) = \partial_{\bq,\ka}v_{\rm C}(\br) + \partial_{\bq,\ka}v_{\rm xc}(\br).
\label{coul-plus-xc}
\end{equation}
The Coulomb contribution $\partial_{\bq,\ka}v_{\rm C}(\br)$ combines the variation of the Hartree and electron-nuclei interactions, which are treated with pseudopotentials.\\
\indent
The average of $\partial_{\bq,\ka}v_{\rm C}(\br)$ over the unit cell volume $\Omega$ is defined as 
\begin{equation}
\Delta_{\ka}(\bq) = \frac{1}{\Omega} \int_{\Omega}\!d\br \,\,\partial_{\bq,\ka}v_{\rm C} (\br).
\label{cell-avg-coul}
\end{equation}
This average is well-behaved at finite and arbitrarily small $\bq$, but the $\bq = 0$ case poses challenges. 
At $\bq\!=\!0$ in \textit{metals}, one can show~\cite{Baroni2001DFPT} that $\Delta_{\ka}(\bq\!=\!0)$ is finite and independent on the direction in which $\bq\!=\!0$ is approached; 
this result is a consequence, loosely speaking, of the fact that electrons in metals redistribute to cancel out the electric field induced by the displacement of the nuclei. 
In insulators (and semiconductors), where this cancellation does not occur, $\Delta_{\ka}(\bq)$ is discontinuous at $\bq\!=\!0$ (the $\Gamma$ point in Fig.~\ref{fig:fig_gconverge}).  
At $\bq\!=\!0$, the current version of \textsc{Quantum Espresso} subtracts from $\partial_{\bq,\ka}v(\br)$ the term $\Delta_{\ka}(\bq\!=\!0)$, regardless of the type of material,  
thus making $\partial_{\bq,\ka}v(\br)$ discontinuous in both metals and insulators. 
In metals, $\Delta_{\ka}(\bq=0)$ is then added back at $\bq\!=\!0$~\cite{Calandra2010Wannier}, so that the perturbation potential $\partial_{\bq,\ka}v(\br)$ stored to disk $-$ and thus, 
the coarse grid e-ph matrix elements [$h_{ij}^{\ka}(\bk,\bq)$ in the AO workflow, and $g_{mn\nu}(\bk,\bq)$ in the WF workflow] $-$ are continuous at $\bq\!=\!0$ for metals. 
In insulators, $\Delta_{\ka}(\bq=0)$ is not added back at $\bq\!=\!0$; the perturbation potential $\partial_{\bq,\ka}v(\br)$ is thus discontinuous and not well-defined at $\bq\!=\!0$, and so are the e-ph matrix elements at $\bq\!=\!0$.\\
\indent 
By contrast, the interpolated e-ph matrix elements are, by construction, continuous functions of $\bq$ near and at $\bq\!=\!0$; this poses no problems in metals, 
but in insulators and semiconductors the interpolation joins continuously e-ph matrix elements at $\bq$-points across the discontinuity, leading to discrepancies between the interpolated and directly computed (with DFT plus DFPT) results.  
For non-zero $\bq$-points inside the region defined by the smallest coarse grid vectors $\bq_{\rm c}$, the e-ph matrix elements are \textit{correct} when computed directly from DFT plus DFPT, 
but only approximate when interpolated. 
The interpolation can thus be improved systematically by using denser coarse $\bq_{\rm c}$ grids, as shown in Fig.~\ref{fig:fig_gconverge}, 
because a larger number of correct e-ph matrix elements near $\bq\!=\!0$ are employed in the interpolation.\\ 
\indent 
On this basis, we analyze the trends in the e-ph matrix elements for silicon and diamond near $\bq\!=\!0$ (the $\Gamma$ point in Fig.~\ref{fig:fig_gconverge}) along the K$-$$\Gamma$ direction, which are highlighted in the inset of Fig.~\ref{fig:fig_gconverge}. 
The discontinuity in the direct DFT plus DFPT calculations at $\Gamma$, which has a value of $\delta_1$ + $\delta_2$, derives from two different sources.
One is the aforementioned discontinuity of the e-ph perturbation potential for insulators, which results in a discontinuity with a value of $\delta_1$ at $\Gamma$. 
The second source is the averaging procedure of the e-ph matrix elements when there are electron and/or phonon degeneracies.  
Silicon exhibits several degeneracies at $\Gamma$ (3-fold for both electrons and phonons), and averaging the e-ph matrix elements 
over degenerate electronic states and phonon modes results in a discontinuity with a value of $\delta_2$ at $\Gamma$. 
This discontinuity is not physical $-$ it simply derives from choosing a particular approach for averaging over degenerate states. 
For example, the interpolated e-ph matrix elements between two neighboring points of a coarse $\bq_{\rm c}$ grid (e.g., the points $[-\frac{1}{8}, \frac{1}{8}, 0]$ and $[0, 0, 0]$ in the inset) are smooth functions of $\bq$, 
so that the sharp discontinuity $\delta_2$ cannot be due to the interpolation procedure, but rather has to be due to averaging over degeneracies.\\  
\indent
In metals, the DFPT discontinuity $\delta_1$ is absent, since the e-ph perturbation potential employed in the calculation is continuous as mentioned above. 
This trend is verified in boron-doped diamond, for which the interpolated e-ph matrix elements are shown in Fig.~S2 of the Supplemental Material. 
As expected, the $\delta_1$ discontinuity is absent; the discontinuity $\delta_2$ due to the degeneracies is still present, but the interpolation overall converges rapidly with respect to the coarse $\bq_{\rm c}$ grids.\\
\indent  
The small error near $\bq=0$ in the e-ph interpolation for semiconductors and insulators does not pose a problem for computing physical observables, 
provided that dense enough coarse $\bq$-point grids are used in DFPT. 
In particular, computations of the e-ph self-energy and transport properties involve integrations of the e-ph matrix elements over the entire BZ; 
small errors in the integrand (i.e., the e-ph matrix elements) over a small BZ region near $\bq \!=\! 0$ cannot affect the integral appreciably, unless the e-ph matrix elements are singular at $\bq\!=\!0$. 
The BZ region affected by the error has a volume of $\Omega_{\rm BZ} / N_{\bq}$, where $N_{\bq}$ is the number of points in the coarse $\bq$-point grid, and $\Omega_{\rm BZ}$ the BZ volume; since $N_{\bq} \approx$~1,000 in a typical calculation, this BZ region is very small.\\  
\indent
Additionally, since the e-ph matrix elements vanish at $\bq\!=\!0$ for acoustic phonons, interpolation errors due to the $\bq\!=\!0$ discontinuity are only relevant for optical phonons, but they do not pose a challenge as noted above unless the e-ph matrix elements are singular at $\bq\!=\!0$. 
Polar materials deserve a separate mention. The e-ph interactions are long-ranged for polar phonons, and the e-ph matrix elements for the longitudinal optical (LO) mode diverge  
at $\bq\!=\!0$ in bulk polar materials.  
The current \textit{ab initio} approach is to interpolate only the short-ranged part of the LO-mode e-ph coupling, and then add an analytical expression in reciprocal space for the LO-mode long-range e-ph coupling, which is dominant near $\bq=0$ and independent of the DFPT e-ph perturbation potential. 
One can accurately reproduce the behavior of the LO e-ph matrix elements near $\bq\!=\!0$, and the singularity can be integrated using dense random grids~\cite{Zhou2016eph_polar}. 
We conclude that the DFPT e-ph perturbation potential at $\bq\!=\!0$ does not pose additional challenges in polar materials. 
%
%
%
%
\section{Discussion}
\vspace{-10pt}
Our results, which establish the accuracy of the AO basis set for computing e-ph interactions in materials, make a compelling case for using a \textit{fixed} localized basis set in e-ph calculations. 
The equations and workflows derived here are general, and can be adapted to arbitrary localized basis sets, including Gaussian-type orbitals (GTOs) commonly employed in quantum chemistry codes~\cite{nwchem}. 
This point is interesting since post-Hartree-Fock \textit{ab initio} methods $-$ e.g., the coupled cluster approach $-$ employing correlated wavefunctions are typically implemented using GTOs~\cite{Garnet-1,Garnet-2}. 
Interfacing these methods with e-ph calculations may enable studies of e-ph interactions in strongly correlated materials.\\
\indent
A fixed basis set such as the AOs employed here has both advantages and disadvantaged compared to WFs. 
WFs are widely used to obtain accurate interpolated band structures, 
but they are a material-specific basis set that needs to be generated through a trial-and-error approach~\cite{Souza2001}. 
By contrast, AOs are not associated with a specific material. They have been traditionally used in quantum chemistry methods, 
and only more recently to accurately interpolate electronic band structures~\cite{Agapito2016TightBinding}.
Since AOs and other fixed localized basis sets are readily available and are not material-specific, they can automate the computation and interpolation of e-ph matrix elements. 
Our AO-based e-ph workflow can be employed in high-throughput calculations and materials discovery studies because, contrary to WFs, there are no challenges in generating the localized basis set.
The AO method is also suitable for studying e-ph interactions and electron dynamics in structurally complex systems, such as surfaces, interfaces, or large unit cells containing defects, for which WFs cannot be readily obtained.\\ 
\indent
Lastly, we point out some drawbacks of the AO basis set. The small deviations between the AO and WF interpolation results derive mainly from the incompleteness of the AO basis set. The latter introduces a small error in the expansion of the DFT electronic wavefunctions, and thus in the e-ph matrix elements in the AO Bloch sum basis (see Eq.~\ref{eq:h-ka-ij}) provided as input in the AO interpolation. 
This error is carried through the workflow into the final interpolated e-ph matrix elements. 
These truncation errors are very small with the double-$\zeta$ polarized basis set employed here, but we have verified that the accuracy of single-$\zeta$ and single-$\zeta$ polarized basis sets is less satisfactory. Increasing the size of the basis set beyond a double-$\zeta$ polarized AO basis will increase the accuracy of the interpolated electronic wavefunctions and e-ph matrix elements. However, larger basis sets significantly increase the computational cost and memory requirements, so that one should seek a tradeoff between accuracy and cost.  
A merit of the WF interpolation is that of reproducing exactly the coarse-grid DFT wavefunctions and e-ph matrix elements at fine-grid points that are also present in the coarse grids. Since the interpolated e-ph matrix elements are smooth in the BZ, this leads to an overall slightly superior accuracy of the WF interpolation method, which constitutes the main advantage of WFs over AOs. An additional advantage of the WFs is that they are a \textit{minimal} localized basis set for a given number of bands of interest. Employing AOs or other localized basis sets results in a larger number of basis functions, and thus larger matrices employed in the e-ph interpolation procedure.
%
%
%
%
\section{Conclusions}
\vspace{-10pt} 
We presented a new method that employs AOs to compute and interpolate the e-ph matrix elements. Benchmark AO calculations of e-ph matrix elements and e-ph self-energies show an accuracy comparable to that of WF interpolation. 
The small deviation between the AO and WF results is due to truncation errors in the AO basis set and the resulting approximate description of the interpolated electron wavefunctions. 
Several benefits of the AO-based e-ph calculations are outlined. Since they are a fixed basis set that can be stored in a database, AOs can automate e-ph calculations, and make them possible for chemically and structurally complex materials. %
%
%
%
%
%
%
%
%
\newpage
\appendix
\vspace{-30pt}
\section{The e-ph perturbation potential}\label{sec:appendixA}
\vspace{-10pt}
Within the Born-Oppenheimer approximation, the potential of the crystal $V(\br;\{\bR\})$ depends parametrically on the positions of the $N$ atoms in the crystal, 
which are given by the $3N$-dimensional vector $\{\bR\} = [\cdots, \bR_{p\ka}, \cdots]$, where $p$ labels the unit cell, $\kappa$ the atom, and $\alpha$ the Cartesian direction. 
At each position $\br$, the Taylor expansion of the crystal potential around the equilibrium positions $\{\bt\}$ is 
\begin{align*}
V(\{\bR\}) &= V(\{\bt\}) + \left( \{\bR\} - \{\bt\} \right) \cdot \nabla V|_{\{\bt\}} \\
           &\quad + \frac{1}{2} \left[ \left(\{\bR\} - \{\bt\} \right) \cdot \nabla\right]^2 V|_{\{\bt\}} + \ldots ,
\end{align*}
where the gradient in the $3N$ dimensional space defined by the atomic positions is $\nabla = [ \cdots, \pdv{}{R_{p \ka}}, \cdots ]$. 
One can define a displacement vector of the atoms from their equilibrium positions, $\{\bu\} = \{\bR\} - \{\bt\} = [\cdots, u_{p\ka}, \cdots]$.\\
\indent
%
A phonon with mode $\nu$ and crystal momentum $\bq$ displaces the atoms (with mass $M_\kappa$) from their equilibrium positions, leading to a displacement 
vector $u^{\nu\bq}_{p\ka}=\frac{1}{\sqrt{M_{\kappa}}}e^{i\bq \cdot \bR_p}e^{\ka}_{\nu\bq}$, where $e^{\ka}_{\nu\bq}$ is the phonon eigenvector. 
Therefore, using the Taylor expansion above, the perturbation potential due to the phonon mode, up to the term linear in the atomic displacements, reads: 
\begin{subequations}
\label{eq:deltaVq-1}
\begin{align} 
\Delta_{\nu\bq} V(\br) &= \sum_{p \ka} u^{\nu \mathbf{q}}_{p \ka} \partial_{\bR_p,\ka} V(\br) \label{eq:deltaVq-1a}\\
&= \sum_{\ka}\frac{1}{\sqrt{M_{\kappa}}}e^{\ka}_{\nu\bq}\partial_{\bq,\ka} V(\br).\label{eq:deltaVq-1b}
\end{align}
\end{subequations}
Here, we introduced the derivative of the potential with respect to a change in the position of an atom (and thus, with respect to its displacement) in a given Cartesian direction, 
and the corresponding Fourier transform,
\begin{subequations}
\label{eq:dV}
\begin{align}
\partial_{\bR_p,\ka} V(\br) &\equiv \left . \pdv{ V(\br;\{\bR\}) }{ R_{p \ka} } \right|_{\{\bt\}} = \left. \pdv{ V(\br;\{\bR\}) }{ u_{p \ka}} \right|_{ \{\bu\}=0 },\label{eq:dV-R}\\
\partial_{\bq,\ka} V(\br)   &= \sum_{\bR_p} e^{i\bq\cdot\bR_p} \partial_{\bR_p,\ka} V(\br).
\label{eq:dV-q}
\end{align}
\end{subequations}
%

%
%
\section{Fourier transform of the AO e-ph matrix elements}\label{sec:appendixB}
\vspace{-10pt}
Using the Bloch sum of AOs defined in Eq.~\ref{eq:bsao}, we derive the double Fourier transform employed in Eq.~\ref{eq:h-doubleft}. 
First, we establish the following result:
\begin{align}
\label{eq:identity-1}
&\mel{\phi_i(\br-\bR_e')}{\dVq}{\phi_j(\br-\bR_e)}\notag\\
&\qquad\qquad = \mel{\phi_i(\br)}
{\partial_{\bq,\ka}V(\br+\bR_e')}
{\phi_j(\br-(\bR_e-\bR_e'))}\notag \\
&\qquad\qquad = e^{i\bq\cdot\bR_e'}\mel{\phi_i(\br)}
{\partial_{\bq,\ka} V(\br)}
{\phi_j(\br-(\bR_e-\bR_e'))},
\end{align}
where in the first line we changed the integration variable using $\br \!\rightarrow\! \br+\bR_e'$. In the last line, we used $\partial_{\bq,\ka} V(\br+\bR_e') = e^{i\bq\cdot\bR_e'}\partial_{\bq,\ka} V(\br)$, which comes from the fact that the perturbation potential $\partial_{\bq,\ka} V(\br) = e^{i\bq\cdot\br}\partial_{\bq,\ka}v(\br)$ can be expressed in terms of the lattice-periodic function $\partial_{\bq,\ka}v(\br+\bR_e') = \partial_{\bq,\ka}v(\br)$.\\
\indent 
Using this result, we write the e-ph matrix element in the AO Bloch sum basis as:
\begin{align}
\label{eq:identity-2}
\!\!\!\!h^{\ka}_{ij}(\bk,\bq)&=
\mel{\Phi_{i\bk+\bq}(\br)}
{\dVq}
{\Phi_{j\bk}(\br)} \notag \\
&=\frac{1}{N_e}
\sum_{\bR_e'}\sum_{\bR_e} e^{i\bk\cdot(\bR_e - \bR_e')}\,e^{-i\bq\cdot\bR_e'} \notag \\
&\quad\times \mel{\phi_i(\br-\bR_e')}
{\dVq}
{\phi_j(\br-\bR_e)} \notag \\
&=\frac{1}{N_e}
\sum_{\bR_e'}\sum_{\bR_e - \bR_e'} e^{i\bk\cdot(\bR_e - \bR_e')} \notag \\
&\quad\times \mel{\phi_i(\br)}
{\dVq}
{\phi_j(\br-(\bR_e-\bR_e'))} \notag \\
&=
\sum_{\bR_e} e^{i\bk\cdot\bR_e}
\mel{\phi_i(\br)}
{\dVq}
{\phi_j(\br-\bR_e)} \notag \\
&=
\sum_{\bR_e,\bR_p} e^{i\bk\cdot\bR_e + i\bq\cdot\bR_p}
\mel{\phi_i(\br)}
{\dVRp}
{\phi_j(\br-\bR_e)} \notag \\
&= \sum_{\bR_e,\bR_p} e^{i(\bk\cdot\bR_e + \bq\cdot\bR_p)}
h^{\ka}_{ij}(\bR_e,\bR_p),
\end{align}
where we used $\dVq = \sum_{\bR_p} e^{i\bq\cdot\bR_p} \dVRp$ (see Eq.~\ref{eq:partial-pertpot}). This double Fourier transform is the result employed in Eq.~\ref{eq:h-doubleft}. 
%

%
%
\section{Using symmetry to compute the perturbation potential in the full BZ}\label{sec:appendixC}
\vspace{-10pt}
\begin{figure}[!htpb] 
\includegraphics[width=8.5cm]{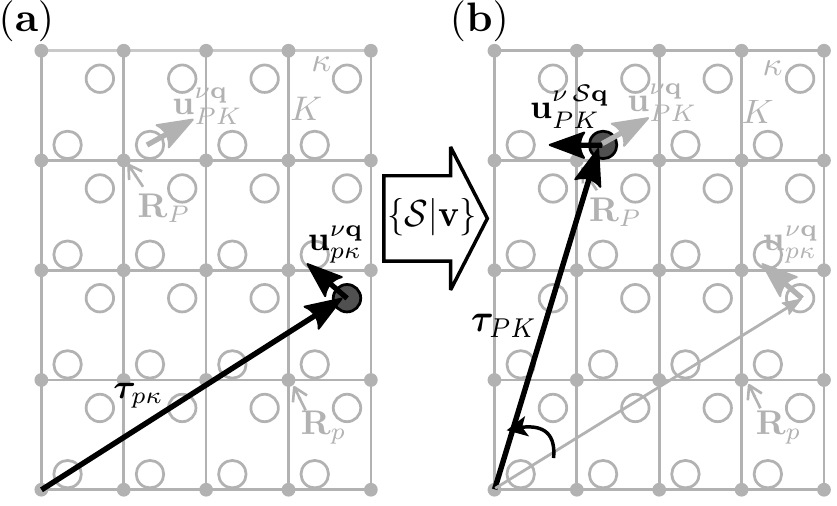}
\caption{(a) Schematic of a phonon with mode index $\nu$ and wavevector $\bq$ frozen in a crystal; only the phonon displacement vectors for the atomic sites $\bt_{\pk}$ and $\bt_{PK}$ are shown for illustration. (b) After applying the symmetry operation $\symop$, the crystal transforms into itself, and the atom at $\bt_{\pk}$ is transformed to the position $\bt_{PK}$.} %
\label{fig:dv_rotation}
\end{figure}
We derive the equation employed to evaluate the perturbation potential in the full BZ starting from calculations in the irreducible wedge. 
The symmetry group of the crystal consists of combinations of point group symmetry operations $\cS$ and fractional translations $\mathbf{v}$. 
These space group symmetry operations, denoted as $\symop$, transform the crystal into itself. 
Our goal is to derive an equation to transform the perturbation potential $\partial_{\bq,\ka} V(\br)$, computed at $\bq$-points in the irreducible wedge, 
to the perturbation potential $\partial_{\Sq,\ka} V(\br)$ computed at points $\cS \bq$ spanning the entire BZ.\\ 
\indent
First, we give the effect of symmetry operations on:\\
\textit{i}) The crystal structure: The symmetry operations transform an atom $\pk$ into the equivalent site $PK$, as seen in Fig.~\ref{fig:dv_rotation}. The new equilibrium atomic position is
\begin{equation}
\symop  \bt_{\pk} = \cS \bt_{\pk} + \bv = \bt_{PK}.
\label{eq:tau-PK}
\end{equation}
The inverse of this transformation is
\begin{equation}
\symop^{-1}  \bt_{\pk} = \cS^{-1} \bt_{\pk} - \cS^{-1} \bv.
\label{eq:tau-PK-inv}
\end{equation}

\textit{ii}) The displacement vector $\bu^{\nu\bq}_{p\kappa}$ of a phonon mode: As shown in Ref.~\onlinecite{Maradudin1968}, 
the transformed displacement vector belongs to the point $\Sq$, and reads:
\begin{equation}
\bu^{\nu\,\Sq}_{PK} = \cS \bu^{\nu\bq}_{\pk}.
\label{eq:u-PK}
\end{equation}

\textit{iii}) A scalar function $g(\br)$:
\begin{align}
\label{eq:symop-g}
\symop g(\br) &= g(\isymop \br)\\
&= g(\cS^{-1}\br-\cS^{-1}\bv )
\end{align}

\textit{iv}) The perturbation potential $\Delta_{\nu\bq} V(\br)$ (see Eq.~\ref{eq:pertpot}):
\begin{align}
\symop  \Delta_{\nu\bq} V(\br) &= \Delta_{\nu\bq} V(\isymop \br) \\
&= \Delta_{\nu\,\Sq} V(\br),  
\label{eq:rot-dvscf}
\end{align}
where the last equality can be derived using the methods in Ref.~\cite{Maradudin1968}. 
We also write the relation between the components of the phonon eigenvector at symmetry-related points $\bq$ and $\Sq$, given in Eq.~2.33 of Ref.~\cite{Maradudin1968}:
\begin{equation}\label{eq:rot-e}
e^{K\alpha}_{\nu\,\Sq} = \sum_{\kappa\beta} \Gamma^{\bq,\symop}_{K\alpha,\kappa\beta} e^{\kappa\beta}_{\nu\bq},
\end{equation}
where the matrix $\Gamma^{\bq,\symop}$ is defined as
\begin{equation}\label{eq:Gamma}
\Gamma^{\bq,\symop}_{K\alpha,\kappa\beta} = e^{i\bq \cdot [\isymop \bt_K - \bt_{\kappa}]} \mathcal{S}_{\alpha\beta}.
\end{equation}
Here, $\bt_K = \bt_{PK} - \bR_P$ and $\bt_\kappa = \bt_{p\kappa} - \bR_p$ are the atomic positions of the $K^{\rm th}$ and $\kappa^{\rm th}$ atoms relative to the origin of the unit cell.

Combining Eqs.~\ref{eq:deltaVq-1b} and \ref{eq:rot-dvscf}, we write 
\begin{subequations}
\label{eq:deltaVq-2}
\begin{align}
\Delta_{\nu\,\Sq} V(\br)
&= \sum_{\ka}\frac{1}{\sqrt{M_{\kappa}}}e^{\ka}_{\nu\bq}\partial_{\bq,\ka}V(\isymop \br)\label{eq:deltaVq-2a}\\
&= \sum_{\ka}\frac{1}{\sqrt{M_{\kappa}}}e^{\ka}_{\nu\,\Sq}\partial_{\Sq,\ka}V(\br).\label{eq:deltaVq-2b}
\end{align}
\end{subequations}
The first line, Eq.~\ref{eq:deltaVq-2a}, can be simplified by rewriting $e^{\ka}_{\nu\bq}$ through the inverse of Eq.~\ref{eq:rot-e}:
\begin{flalign}\label{eq:rot-dvscf-1}
&\Delta_{\nu\,\Sq}V(\br) = \sum_{\ka} \frac{1}{\sqrt{M_{\kappa}}} e^{\ka}_{\nu\bq}
\partial_{\bq,\ka} V(\isymop \br)& \notag
\\
&= \sum_{\ka} \frac{1}{\sqrt{M_{\kappa}}} 
\sum_{K\beta} e^{-i\bq \cdot [\isymop \bt_K - \bt_{\kappa}]} [\cS^{-1}]_{\alpha\beta}\, e^{K\beta}_{\nu\,\Sq} &\notag
\\
&\quad \times \partial_{\bq,\ka} V(\isymop \br)&\notag
\\
&= \sum_{K\beta} \frac{1}{\sqrt{M_{K}}} e^{K\beta}_{\nu\,\Sq} \sum_{\ka} \Big[ e^{i\bq \cdot \cS^{-1}\bv}e^{i\bq \cdot \bt_{\kappa} -i\bq \cdot \cS^{-1}\bt_K} &\notag
\\
&\quad \times [\cS^{-1}]_{\alpha\beta}\,\, \partial_{\bq,\kappa\alpha} V(\isymop \br) \Big] &\notag
\\
&= \sum_{\ka} \frac{1}{\sqrt{M_{\kappa}}} e^{\ka}_{\nu\,\Sq} \Big[ e^{i \Sq \cdot \bv } \sum_{\kappa' \beta} e^{i \bq \cdot \bt_{\kappa'} - i \Sq \cdot \bt_{\kappa} } \notag&\\
&\quad \times [\cS^{-1}]_{\beta\alpha}\,\, \partial_{\bq,\kappa'\beta} V(\isymop \br) \Big].&
\end{flalign}
In the second equality, we used Eq.~\ref{eq:tau-PK-inv} as well as $M_K\!=\!M_{\kappa}$ since symmetry-equivalent atoms belong to the same species; in the last equality, 
we changed the variables from $K\beta \rightarrow \kappa\alpha$ and from $\ka \rightarrow \kappa'\beta$. 
By comparing \ref{eq:deltaVq-2b} with the last line of Eq.~\ref{eq:rot-dvscf-1}, we find
\begin{align}\label{eq:rot-dvscf-2}
\partial_{\Sq,\ka} V(\br) &=  e^{i \Sq \cdot \bv } \sum_{\kappa'\beta} e^{i \bq \cdot \bt_{\kappa'} - i \Sq \cdot \bt_{\kappa} } \notag \\
&\quad \times [\cS^{-1}]_{\beta\alpha}\,\, \partial_{\bq,\kappa'\beta}V(\isymop \br).
\end{align}

Finally, the result can be expressed in terms of the lattice-periodic e-ph perturbation potential $\partial_{\bq,\ka}v(\br) = e^{-i\bq \cdot \br}\, \partial_{\bq,\ka}V(\br)$ (see Eq.~\ref{periodic-eph-pert}), which is computed directly and stored to disk in \textsc{Quantum Espresso}. 
We thus obtain the final result employed in our calculations: 
\begin{equation}
\partial_{\Sq,\ka}v(\br) = \sum_{\kappa'\beta} e^{i \bq \cdot \bt_{\kappa'} - i \Sq \cdot \bt_{\kappa} } [\cS^{-1}]_{\beta\alpha}
\, \partial_{\bq,\kappa'\beta}v(\isymop \br)
\label{eq:rot-dvscf-3}.
\end{equation}

Note that while the results of this section assume a local potential $V(\br)$, modern DFT implementations use pseudopotentials that introduce a non-local part in the Kohn-Sham potential. 
The non-local contribution to the e-ph perturbation potential can be computed efficiently, with the analytical formula given in Eq.~A14 of Ref.~\cite{Baroni2001DFPT}, directly for all the $\bq$ points in the full BZ, without resorting to symmetry operations. Therefore, symmetry is only employed to reduce the computational cost of obtaining the local part of the e-ph perturbation potential.\\
\indent
The dynamical matrices are also evaluated in the full BZ starting from calculations in the irreducible wedge, using the following expression \cite{Maradudin1968}:
\begin{equation}
D(\Sq) = \Gamma^{\bq,\symop} D(\bq) [\Gamma^{\bq,\symop}]^\dagger.
\end{equation}
%
%
\section{Hamiltonian in AO Bloch sum basis}\label{sec:appendixD}
\vspace{-10pt}
We derive the relationship used in Eq.~\ref{eq:H-interp} between the real-space AO Hamiltonian matrix $H(\bR)$ and the reciprocal-space Hamiltonian matrix $H(\bk)$ in the AO Bloch sum basis. 
Using Bloch sums defined in Eq.~\ref{eq:bsao}, we write:
\begin{flalign*}
\begin{split}
H_{ij}(\bk)&=
\mel{\Phi_{i\bk}(\br)} {\hat{H}(\br)} {\Phi_{j\bk}(\br)} \\
&=\frac{1}{N_e}
\sum_{\bR_e'}\sum_{\bR_e} e^{i\bk\cdot(\bR_e - \bR_e')} \\
&\quad\times \mel{\phi_i(\br-\bR_e')} {\hat{H}(\br)} {\phi_j(\br-\bR_e)}\\
&=\frac{1}{N_e}
\sum_{\bR_e'}\sum_{\bR_e } e^{i\bk\cdot(\bR_e - \bR_e')} \\
&\quad\times \mel{\phi_i(\br)}{\hat{H}(\br)}{\phi_j(\br-(\bR_e-\bR_e'))}\\
&=
\sum_{\bR_e} e^{i\bk\cdot\bR_e}
\mel{\phi_i(\br)} {\hat{H}(\br)} {\phi_j(\br-\bR_e)}\\
&= \sum_{\bR_e} e^{i\bk\cdot\bR_e} H_{ij}(\bR_e).
\end{split}
\end{flalign*}

%
%
\section{A note on Wannier function interpolation}\label{sec:mlwf}
\vspace{-10pt}
We derive the result quoted in Eq.~\ref{eq:wfhij}. Expanding the Bloch states in terms of Bloch sums of WFs, the e-ph matrix element can be written as
\begin{flalign*}
\begin{split}
g^{\ka}_{mn}(\bk,\bq)&=
\mel{\psi_{m\bk+\bq}(\br)} {\dVq} {\psi_{n\bk}(\br)} \\
&=\sum_{ij}U^{\bk+\bq}_{mi}\, (U^{\bk}_{nj})^* \mel{W_{i\bk+\bq}(\br)}{\dVq}{W_{j\bk}(\br)} \\
&=\sum_{ij}U^{\bk+\bq}_{mi}\left( U^{\bk}_{nj} \right)^* h^{\ka}_{ij}(\bk,\bq).
\end{split}
\end{flalign*}
In matrix form, this linear transformation and its inverse, which is employed in Eq.~\ref{eq:wfhij}, read respectively: 
\begin{subequations}
\begin{align}
g^{\ka}(\bk,\bq) &= U^{\bk+\bq}\, h^{\ka}(\bk,\bq) \, \left( U^{\bk} \right)^{\dagger} \label{eq:g-ka-wf}\\
h^{\ka}(\bk,\bq) &= \left(U^{\bk+\bq}\right)^{\dagger}\, g^{\ka}(\bk,\bq) \, U^{\bk}\label{eq:h-ka-wf}. 
\end{align}
\end{subequations}

\acknowledgements
\vspace{-10pt}
L.A. thanks Prof. M. Wierzbowska and Dr. A. Ferretti for technical discussions. This work was supported by the National Science Foundation under grant no. SI2-SSE-1642443. This  research used  resources  of  the  National  Energy  Research  Scientific Computing  Center,  a  DOE  Office  of  Science  User  Facility supported by the Office of Science of the U.S. Department of Energy under Contract No. DE-AC02-05CH11231. We also acknowledge the Extreme Science and Engineering Discovery Environment (XSEDE) for providing high-performance computing resources and technical assistance.


\begin{thebibliography}{37}%
\makeatletter
\providecommand \@ifxundefined [1]{%
 \@ifx{#1\undefined}
}%
\providecommand \@ifnum [1]{%
 \ifnum #1\expandafter \@firstoftwo
 \else \expandafter \@secondoftwo
 \fi
}%
\providecommand \@ifx [1]{%
 \ifx #1\expandafter \@firstoftwo
 \else \expandafter \@secondoftwo
 \fi
}%
\providecommand \natexlab [1]{#1}%
\providecommand \enquote  [1]{``#1''}%
\providecommand \bibnamefont  [1]{#1}%
\providecommand \bibfnamefont [1]{#1}%
\providecommand \citenamefont [1]{#1}%
\providecommand \href@noop [0]{\@secondoftwo}%
\providecommand \href [0]{\begingroup \@sanitize@url \@href}%
\providecommand \@href[1]{\@@startlink{#1}\@@href}%
\providecommand \@@href[1]{\endgroup#1\@@endlink}%
\providecommand \@sanitize@url [0]{\catcode `\\12\catcode `\$12\catcode
  `\&12\catcode `\#12\catcode `\^12\catcode `\_12\catcode `\%12\relax}%
\providecommand \@@startlink[1]{}%
\providecommand \@@endlink[0]{}%
\providecommand \url  [0]{\begingroup\@sanitize@url \@url }%
\providecommand \@url [1]{\endgroup\@href {#1}{\urlprefix }}%
\providecommand \urlprefix  [0]{URL }%
\providecommand \Eprint [0]{\href }%
\providecommand \doibase [0]{http://dx.doi.org/}%
\providecommand \selectlanguage [0]{\@gobble}%
\providecommand \bibinfo  [0]{\@secondoftwo}%
\providecommand \bibfield  [0]{\@secondoftwo}%
\providecommand \translation [1]{[#1]}%
\providecommand \BibitemOpen [0]{}%
\providecommand \bibitemStop [0]{}%
\providecommand \bibitemNoStop [0]{.\EOS\space}%
\providecommand \EOS [0]{\spacefactor3000\relax}%
\providecommand \BibitemShut  [1]{\csname bibitem#1\endcsname}%
\let\auto@bib@innerbib\@empty
\bibitem [{\citenamefont {Bernardi}(2016)}]{Bernardi-review}%
  \BibitemOpen
  \bibfield  {author} {\bibinfo {author} {\bibfnamefont {M.}~\bibnamefont
  {Bernardi}},\ }\href {\doibase 10.1140/epjb/e2016-70399-4} {\bibfield
  {journal} {\bibinfo  {journal} {Eur. Phys. J. B}\ }\textbf {\bibinfo {volume}
  {89}},\ \bibinfo {pages} {239} (\bibinfo {year} {2016})}\BibitemShut
  {NoStop}%
\bibitem [{\citenamefont {Giustino}(2017)}]{Giustino2017}%
  \BibitemOpen
  \bibfield  {author} {\bibinfo {author} {\bibfnamefont {F.}~\bibnamefont
  {Giustino}},\ }\href {\doibase 10.1103/RevModPhys.89.015003} {\bibfield
  {journal} {\bibinfo  {journal} {Rev. Mod. Phys.}\ }\textbf {\bibinfo {volume}
  {89}},\ \bibinfo {pages} {015003} (\bibinfo {year} {2017})}\BibitemShut
  {NoStop}%
\bibitem [{\citenamefont {Gonze}\ and\ \citenamefont
  {Lee}(1997)}]{Gonze1997_DFPT}%
  \BibitemOpen
  \bibfield  {author} {\bibinfo {author} {\bibfnamefont {X.}~\bibnamefont
  {Gonze}}\ and\ \bibinfo {author} {\bibfnamefont {C.}~\bibnamefont {Lee}},\
  }\href {\doibase 10.1103/PhysRevB.55.10355} {\bibfield  {journal} {\bibinfo
  {journal} {Phys. Rev. B}\ }\textbf {\bibinfo {volume} {55}},\ \bibinfo
  {pages} {10355} (\bibinfo {year} {1997})}\BibitemShut {NoStop}%
\bibitem [{\citenamefont {Baroni}\ \emph {et~al.}(2001)\citenamefont {Baroni},
  \citenamefont {de~Gironcoli}, \citenamefont {Dal~Corso},\ and\ \citenamefont
  {Giannozzi}}]{Baroni2001DFPT}%
  \BibitemOpen
  \bibfield  {author} {\bibinfo {author} {\bibfnamefont {S.}~\bibnamefont
  {Baroni}}, \bibinfo {author} {\bibfnamefont {S.}~\bibnamefont
  {de~Gironcoli}}, \bibinfo {author} {\bibfnamefont {A.}~\bibnamefont
  {Dal~Corso}}, \ and\ \bibinfo {author} {\bibfnamefont {P.}~\bibnamefont
  {Giannozzi}},\ }\href {\doibase 10.1103/RevModPhys.73.515} {\bibfield
  {journal} {\bibinfo  {journal} {Rev. Mod. Phys.}\ }\textbf {\bibinfo {volume}
  {73}},\ \bibinfo {pages} {515} (\bibinfo {year} {2001})}\BibitemShut
  {NoStop}%
\bibitem [{\citenamefont {Savrasov}\ \emph {et~al.}(1994)\citenamefont
  {Savrasov}, \citenamefont {Savrasov},\ and\ \citenamefont
  {Andersen}}]{Savrasov1994eph}%
  \BibitemOpen
  \bibfield  {author} {\bibinfo {author} {\bibfnamefont {S.~Y.}\ \bibnamefont
  {Savrasov}}, \bibinfo {author} {\bibfnamefont {D.~Y.}\ \bibnamefont
  {Savrasov}}, \ and\ \bibinfo {author} {\bibfnamefont {O.~K.}\ \bibnamefont
  {Andersen}},\ }\href {\doibase 10.1103/PhysRevLett.72.372} {\bibfield
  {journal} {\bibinfo  {journal} {Phys. Rev. Lett.}\ }\textbf {\bibinfo
  {volume} {72}},\ \bibinfo {pages} {372} (\bibinfo {year} {1994})}\BibitemShut
  {NoStop}%
\bibitem [{\citenamefont {Savrasov}\ and\ \citenamefont
  {Savrasov}(1996)}]{Savrasov1996eph}%
  \BibitemOpen
  \bibfield  {author} {\bibinfo {author} {\bibfnamefont {S.~Y.}\ \bibnamefont
  {Savrasov}}\ and\ \bibinfo {author} {\bibfnamefont {D.~Y.}\ \bibnamefont
  {Savrasov}},\ }\href {\doibase 10.1103/PhysRevB.54.16487} {\bibfield
  {journal} {\bibinfo  {journal} {Phys. Rev. B}\ }\textbf {\bibinfo {volume}
  {54}},\ \bibinfo {pages} {16487} (\bibinfo {year} {1996})}\BibitemShut
  {NoStop}%
\bibitem [{\citenamefont {Deinzer}\ \emph {et~al.}(2003)\citenamefont
  {Deinzer}, \citenamefont {Birner},\ and\ \citenamefont {Strauch}}]{Deinzer}%
  \BibitemOpen
  \bibfield  {author} {\bibinfo {author} {\bibfnamefont {G.}~\bibnamefont
  {Deinzer}}, \bibinfo {author} {\bibfnamefont {G.}~\bibnamefont {Birner}}, \
  and\ \bibinfo {author} {\bibfnamefont {D.}~\bibnamefont {Strauch}},\ }\href
  {\doibase 10.1103/PhysRevB.67.144304} {\bibfield  {journal} {\bibinfo
  {journal} {Phys. Rev. B}\ }\textbf {\bibinfo {volume} {67}},\ \bibinfo
  {pages} {144304} (\bibinfo {year} {2003})}\BibitemShut {NoStop}%
\bibitem [{\citenamefont {Zhou}\ and\ \citenamefont
  {Bernardi}(2016)}]{Zhou2016eph_polar}%
  \BibitemOpen
  \bibfield  {author} {\bibinfo {author} {\bibfnamefont {J.-J.}\ \bibnamefont
  {Zhou}}\ and\ \bibinfo {author} {\bibfnamefont {M.}~\bibnamefont
  {Bernardi}},\ }\href {\doibase 10.1103/PhysRevB.94.201201} {\bibfield
  {journal} {\bibinfo  {journal} {Phys. Rev. B}\ }\textbf {\bibinfo {volume}
  {94}},\ \bibinfo {pages} {201201} (\bibinfo {year} {2016})}\BibitemShut
  {NoStop}%
\bibitem [{\citenamefont {Mustafa}\ \emph {et~al.}(2016)\citenamefont
  {Mustafa}, \citenamefont {Bernardi}, \citenamefont {Neaton},\ and\
  \citenamefont {Louie}}]{Bernardi-noble}%
  \BibitemOpen
  \bibfield  {author} {\bibinfo {author} {\bibfnamefont {J.~I.}\ \bibnamefont
  {Mustafa}}, \bibinfo {author} {\bibfnamefont {M.}~\bibnamefont {Bernardi}},
  \bibinfo {author} {\bibfnamefont {J.~B.}\ \bibnamefont {Neaton}}, \ and\
  \bibinfo {author} {\bibfnamefont {S.~G.}\ \bibnamefont {Louie}},\ }\href
  {\doibase 10.1103/PhysRevB.94.155105} {\bibfield  {journal} {\bibinfo
  {journal} {Phys. Rev. B}\ }\textbf {\bibinfo {volume} {94}},\ \bibinfo
  {pages} {155105} (\bibinfo {year} {2016})}\BibitemShut {NoStop}%
\bibitem [{\citenamefont {Bernardi}\ \emph {et~al.}(2014)\citenamefont
  {Bernardi}, \citenamefont {Vigil-Fowler}, \citenamefont {Lischner},
  \citenamefont {Neaton},\ and\ \citenamefont {Louie}}]{Bernardi2014Si}%
  \BibitemOpen
  \bibfield  {author} {\bibinfo {author} {\bibfnamefont {M.}~\bibnamefont
  {Bernardi}}, \bibinfo {author} {\bibfnamefont {D.}~\bibnamefont
  {Vigil-Fowler}}, \bibinfo {author} {\bibfnamefont {J.}~\bibnamefont
  {Lischner}}, \bibinfo {author} {\bibfnamefont {J.~B.}\ \bibnamefont
  {Neaton}}, \ and\ \bibinfo {author} {\bibfnamefont {S.~G.}\ \bibnamefont
  {Louie}},\ }\href {\doibase 10.1103/PhysRevLett.112.257402} {\bibfield
  {journal} {\bibinfo  {journal} {Phys. Rev. Lett.}\ }\textbf {\bibinfo
  {volume} {112}},\ \bibinfo {pages} {257402} (\bibinfo {year}
  {2014})}\BibitemShut {NoStop}%
\bibitem [{\citenamefont {Bernardi}\ \emph
  {et~al.}(2015{\natexlab{a}})\citenamefont {Bernardi}, \citenamefont
  {Vigil-Fowler}, \citenamefont {Ong}, \citenamefont {Neaton},\ and\
  \citenamefont {Louie}}]{Bernardi2015GaAs}%
  \BibitemOpen
  \bibfield  {author} {\bibinfo {author} {\bibfnamefont {M.}~\bibnamefont
  {Bernardi}}, \bibinfo {author} {\bibfnamefont {D.}~\bibnamefont
  {Vigil-Fowler}}, \bibinfo {author} {\bibfnamefont {C.~S.}\ \bibnamefont
  {Ong}}, \bibinfo {author} {\bibfnamefont {J.~B.}\ \bibnamefont {Neaton}}, \
  and\ \bibinfo {author} {\bibfnamefont {S.~G.}\ \bibnamefont {Louie}},\ }\href
  {\doibase 10.1073/pnas.1419446112} {\bibfield  {journal} {\bibinfo  {journal}
  {Proc. Natl. Acad. Sci. U.S.A.}\ }\textbf {\bibinfo {volume} {112}},\
  \bibinfo {pages} {5291} (\bibinfo {year} {2015}{\natexlab{a}})}\BibitemShut
  {NoStop}%
\bibitem [{\citenamefont {Giustino}\ \emph {et~al.}(2007)\citenamefont
  {Giustino}, \citenamefont {Cohen},\ and\ \citenamefont
  {Louie}}]{Giustino2007EPW}%
  \BibitemOpen
  \bibfield  {author} {\bibinfo {author} {\bibfnamefont {F.}~\bibnamefont
  {Giustino}}, \bibinfo {author} {\bibfnamefont {M.~L.}\ \bibnamefont {Cohen}},
  \ and\ \bibinfo {author} {\bibfnamefont {S.~G.}\ \bibnamefont {Louie}},\
  }\href {\doibase 10.1103/PhysRevB.76.165108} {\bibfield  {journal} {\bibinfo
  {journal} {Phys. Rev. B}\ }\textbf {\bibinfo {volume} {76}},\ \bibinfo
  {pages} {165108} (\bibinfo {year} {2007})}\BibitemShut {NoStop}%
\bibitem [{\citenamefont {Calandra}\ \emph {et~al.}(2010)\citenamefont
  {Calandra}, \citenamefont {Profeta},\ and\ \citenamefont
  {Mauri}}]{Calandra2010Wannier}%
  \BibitemOpen
  \bibfield  {author} {\bibinfo {author} {\bibfnamefont {M.}~\bibnamefont
  {Calandra}}, \bibinfo {author} {\bibfnamefont {G.}~\bibnamefont {Profeta}}, \
  and\ \bibinfo {author} {\bibfnamefont {F.}~\bibnamefont {Mauri}},\ }\href
  {\doibase 10.1103/PhysRevB.82.165111} {\bibfield  {journal} {\bibinfo
  {journal} {Phys. Rev. B}\ }\textbf {\bibinfo {volume} {82}},\ \bibinfo
  {pages} {165111} (\bibinfo {year} {2010})}\BibitemShut {NoStop}%
\bibitem [{\citenamefont {Marzari}\ \emph {et~al.}(2012)\citenamefont
  {Marzari}, \citenamefont {Mostofi}, \citenamefont {Yates}, \citenamefont
  {Souza},\ and\ \citenamefont {Vanderbilt}}]{Marzari2012MLWF}%
  \BibitemOpen
  \bibfield  {author} {\bibinfo {author} {\bibfnamefont {N.}~\bibnamefont
  {Marzari}}, \bibinfo {author} {\bibfnamefont {A.~A.}\ \bibnamefont
  {Mostofi}}, \bibinfo {author} {\bibfnamefont {J.~R.}\ \bibnamefont {Yates}},
  \bibinfo {author} {\bibfnamefont {I.}~\bibnamefont {Souza}}, \ and\ \bibinfo
  {author} {\bibfnamefont {D.}~\bibnamefont {Vanderbilt}},\ }\href {\doibase
  10.1103/RevModPhys.84.1419} {\bibfield  {journal} {\bibinfo  {journal} {Rev.
  Mod. Phys.}\ }\textbf {\bibinfo {volume} {84}},\ \bibinfo {pages} {1419}
  (\bibinfo {year} {2012})}\BibitemShut {NoStop}%
\bibitem [{\citenamefont {Jhalani}\ \emph {et~al.}(2017)\citenamefont
  {Jhalani}, \citenamefont {Zhou},\ and\ \citenamefont
  {Bernardi}}]{Jhalani2017GaN}%
  \BibitemOpen
  \bibfield  {author} {\bibinfo {author} {\bibfnamefont {V.~A.}\ \bibnamefont
  {Jhalani}}, \bibinfo {author} {\bibfnamefont {J.-J.}\ \bibnamefont {Zhou}}, \
  and\ \bibinfo {author} {\bibfnamefont {M.}~\bibnamefont {Bernardi}},\ }\href
  {\doibase 10.1021/acs.nanolett.7b02212} {\bibfield  {journal} {\bibinfo
  {journal} {Nano Lett.}\ }\textbf {\bibinfo {volume} {17}},\ \bibinfo {pages}
  {5012} (\bibinfo {year} {2017})}\BibitemShut {NoStop}%
\bibitem [{\citenamefont {Bernardi}\ \emph
  {et~al.}(2015{\natexlab{b}})\citenamefont {Bernardi}, \citenamefont
  {Mustafa}, \citenamefont {Neaton},\ and\ \citenamefont
  {Louie}}]{Bernardi2015Metals}%
  \BibitemOpen
  \bibfield  {author} {\bibinfo {author} {\bibfnamefont {M.}~\bibnamefont
  {Bernardi}}, \bibinfo {author} {\bibfnamefont {J.}~\bibnamefont {Mustafa}},
  \bibinfo {author} {\bibfnamefont {J.~B.}\ \bibnamefont {Neaton}}, \ and\
  \bibinfo {author} {\bibfnamefont {S.~G.}\ \bibnamefont {Louie}},\ }\href
  {http://dx.doi.org/10.1038/ncomms8044} {\bibfield  {journal} {\bibinfo
  {journal} {Nat. Commun.}\ }\textbf {\bibinfo {volume} {6}},\ \bibinfo {pages}
  {7044} (\bibinfo {year} {2015}{\natexlab{b}})}\BibitemShut {NoStop}%
\bibitem [{\citenamefont {Li}(2015)}]{Li2015eph_converge}%
  \BibitemOpen
  \bibfield  {author} {\bibinfo {author} {\bibfnamefont {W.}~\bibnamefont
  {Li}},\ }\href {\doibase 10.1103/PhysRevB.92.075405} {\bibfield  {journal}
  {\bibinfo  {journal} {Phys. Rev. B}\ }\textbf {\bibinfo {volume} {92}},\
  \bibinfo {pages} {075405} (\bibinfo {year} {2015})}\BibitemShut {NoStop}%
\bibitem [{\citenamefont {{Lee}}\ \emph {et~al.}()\citenamefont {{Lee}},
  \citenamefont {{Zhou}}, \citenamefont {{Agapito}},\ and\ \citenamefont
  {{Bernardi}}}]{Nienen-2018}%
  \BibitemOpen
  \bibfield  {author} {\bibinfo {author} {\bibfnamefont {N.-E.}\ \bibnamefont
  {{Lee}}}, \bibinfo {author} {\bibfnamefont {J.-J.}\ \bibnamefont {{Zhou}}},
  \bibinfo {author} {\bibfnamefont {L.~A.}\ \bibnamefont {{Agapito}}}, \ and\
  \bibinfo {author} {\bibfnamefont {M.}~\bibnamefont {{Bernardi}}},\ }\href
  {https://arxiv.org/abs/1712.00490} {\ }\Eprint
  {http://arxiv.org/abs/arXiv:1712.00490} {arXiv:1712.00490} \BibitemShut
  {NoStop}%
\bibitem [{\citenamefont {Agapito}\ \emph {et~al.}(2013)\citenamefont
  {Agapito}, \citenamefont {Ferretti}, \citenamefont {Calzolari}, \citenamefont
  {Curtarolo},\ and\ \citenamefont
  {Buongiorno~Nardelli}}]{Agapito_2013_projectionsPRB}%
  \BibitemOpen
  \bibfield  {author} {\bibinfo {author} {\bibfnamefont {L.~A.}\ \bibnamefont
  {Agapito}}, \bibinfo {author} {\bibfnamefont {A.}~\bibnamefont {Ferretti}},
  \bibinfo {author} {\bibfnamefont {A.}~\bibnamefont {Calzolari}}, \bibinfo
  {author} {\bibfnamefont {S.}~\bibnamefont {Curtarolo}}, \ and\ \bibinfo
  {author} {\bibfnamefont {M.}~\bibnamefont {Buongiorno~Nardelli}},\ }\href
  {\doibase 10.1103/PhysRevB.88.165127} {\bibfield  {journal} {\bibinfo
  {journal} {Phys. Rev. B}\ }\textbf {\bibinfo {volume} {88}},\ \bibinfo
  {pages} {165127} (\bibinfo {year} {2013})}\BibitemShut {NoStop}%
\bibitem [{\citenamefont {Agapito}\ \emph {et~al.}(2016)\citenamefont
  {Agapito}, \citenamefont {Ismail-Beigi}, \citenamefont {Curtarolo},
  \citenamefont {Fornari},\ and\ \citenamefont
  {Nardelli}}]{Agapito2016TightBinding}%
  \BibitemOpen
  \bibfield  {author} {\bibinfo {author} {\bibfnamefont {L.~A.}\ \bibnamefont
  {Agapito}}, \bibinfo {author} {\bibfnamefont {S.}~\bibnamefont
  {Ismail-Beigi}}, \bibinfo {author} {\bibfnamefont {S.}~\bibnamefont
  {Curtarolo}}, \bibinfo {author} {\bibfnamefont {M.}~\bibnamefont {Fornari}},
  \ and\ \bibinfo {author} {\bibfnamefont {M.~B.}\ \bibnamefont {Nardelli}},\
  }\href {\doibase 10.1103/PhysRevB.93.035104} {\bibfield  {journal} {\bibinfo
  {journal} {Phys. Rev. B}\ }\textbf {\bibinfo {volume} {93}},\ \bibinfo
  {pages} {035104} (\bibinfo {year} {2016})}\BibitemShut {NoStop}%
\bibitem [{\citenamefont {McClain}\ \emph {et~al.}(2017)\citenamefont
  {McClain}, \citenamefont {Sun}, \citenamefont {Chan},\ and\ \citenamefont
  {Berkelbach}}]{Garnet-1}%
  \BibitemOpen
  \bibfield  {author} {\bibinfo {author} {\bibfnamefont {J.}~\bibnamefont
  {McClain}}, \bibinfo {author} {\bibfnamefont {Q.}~\bibnamefont {Sun}},
  \bibinfo {author} {\bibfnamefont {G.~K.-L.}\ \bibnamefont {Chan}}, \ and\
  \bibinfo {author} {\bibfnamefont {T.~C.}\ \bibnamefont {Berkelbach}},\ }\href
  {\doibase 10.1021/acs.jctc.7b00049} {\bibfield  {journal} {\bibinfo
  {journal} {J. Chem. Theory Comput.}\ }\textbf {\bibinfo {volume} {13}},\
  \bibinfo {pages} {1209} (\bibinfo {year} {2017})}\BibitemShut {NoStop}%
\bibitem [{\citenamefont {Booth}\ \emph {et~al.}(2016)\citenamefont {Booth},
  \citenamefont {Tsatsoulis}, \citenamefont {Chan},\ and\ \citenamefont
  {Gr{\"u}neis}}]{Garnet-2}%
  \BibitemOpen
  \bibfield  {author} {\bibinfo {author} {\bibfnamefont {G.~H.}\ \bibnamefont
  {Booth}}, \bibinfo {author} {\bibfnamefont {T.}~\bibnamefont {Tsatsoulis}},
  \bibinfo {author} {\bibfnamefont {G.~K.-L.}\ \bibnamefont {Chan}}, \ and\
  \bibinfo {author} {\bibfnamefont {A.}~\bibnamefont {Gr{\"u}neis}},\ }\href
  {\doibase 10.1063/1.4961301} {\bibfield  {journal} {\bibinfo  {journal} {J.
  Chem. Phys.}\ }\textbf {\bibinfo {volume} {145}},\ \bibinfo {pages} {084111}
  (\bibinfo {year} {2016})}\BibitemShut {NoStop}%
\bibitem [{Note1()}]{Note1}%
  \BibitemOpen
  \bibinfo {note} {Although the Kohn-Sham potential is non-local when using
  pseudopotentials, we will use the simplified notation $V(\protect \mathbf
  {r})$ to denote it, and warn the reader about the role of the non-local part
  of the potential when relevant.}\BibitemShut {Stop}%
\bibitem [{Note2()}]{Note2}%
  \BibitemOpen
  \bibinfo {note} {\protect \url {http://perturbo.caltech.edu/}}\BibitemShut
  {NoStop}%
\bibitem [{\citenamefont {Mostofi}\ \emph {et~al.}(2014)\citenamefont
  {Mostofi}, \citenamefont {Yates}, \citenamefont {Pizzi}, \citenamefont {Lee},
  \citenamefont {Souza}, \citenamefont {Vanderbilt},\ and\ \citenamefont
  {Marzari}}]{Mostofi2014W90}%
  \BibitemOpen
  \bibfield  {author} {\bibinfo {author} {\bibfnamefont {A.~A.}\ \bibnamefont
  {Mostofi}}, \bibinfo {author} {\bibfnamefont {J.~R.}\ \bibnamefont {Yates}},
  \bibinfo {author} {\bibfnamefont {G.}~\bibnamefont {Pizzi}}, \bibinfo
  {author} {\bibfnamefont {Y.-S.}\ \bibnamefont {Lee}}, \bibinfo {author}
  {\bibfnamefont {I.}~\bibnamefont {Souza}}, \bibinfo {author} {\bibfnamefont
  {D.}~\bibnamefont {Vanderbilt}}, \ and\ \bibinfo {author} {\bibfnamefont
  {N.}~\bibnamefont {Marzari}},\ }\href {\doibase
  https://doi.org/10.1016/j.cpc.2014.05.003} {\bibfield  {journal} {\bibinfo
  {journal} {Comput. Phys. Commun.}\ }\textbf {\bibinfo {volume} {185}},\
  \bibinfo {pages} {2309 } (\bibinfo {year} {2014})}\BibitemShut {NoStop}%
\bibitem [{\citenamefont {Giannozzi}\ \emph {et~al.}(2009)\citenamefont
  {Giannozzi}, \citenamefont {Baroni}, \citenamefont {Bonini}, \citenamefont
  {Calandra}, \citenamefont {Car}, \citenamefont {Cavazzoni}, \citenamefont
  {Ceresoli}, \citenamefont {Chiarotti}, \citenamefont {Cococcioni},
  \citenamefont {Dabo}, \citenamefont {{Dal Corso}}, \citenamefont
  {de~Gironcoli}, \citenamefont {Fabris}, \citenamefont {Fratesi},
  \citenamefont {Gebauer}, \citenamefont {Gerstmann}, \citenamefont
  {Gougoussis}, \citenamefont {Kokalj}, \citenamefont {Lazzeri}, \citenamefont
  {Martin-Samos}, \citenamefont {Marzari}, \citenamefont {Mauri}, \citenamefont
  {Mazzarello}, \citenamefont {Paolini}, \citenamefont {Pasquarello},
  \citenamefont {Paulatto}, \citenamefont {Sbraccia}, \citenamefont {Scandolo},
  \citenamefont {Sclauzero}, \citenamefont {Seitsonen}, \citenamefont
  {Smogunov}, \citenamefont {Umari},\ and\ \citenamefont
  {Wentzcovitch}}]{Giannozzi2009}%
  \BibitemOpen
  \bibfield  {author} {\bibinfo {author} {\bibfnamefont {P.}~\bibnamefont
  {Giannozzi}}, \bibinfo {author} {\bibfnamefont {S.}~\bibnamefont {Baroni}},
  \bibinfo {author} {\bibfnamefont {N.}~\bibnamefont {Bonini}}, \bibinfo
  {author} {\bibfnamefont {M.}~\bibnamefont {Calandra}}, \bibinfo {author}
  {\bibfnamefont {R.}~\bibnamefont {Car}}, \bibinfo {author} {\bibfnamefont
  {C.}~\bibnamefont {Cavazzoni}}, \bibinfo {author} {\bibfnamefont
  {D.}~\bibnamefont {Ceresoli}}, \bibinfo {author} {\bibfnamefont {G.~L.}\
  \bibnamefont {Chiarotti}}, \bibinfo {author} {\bibfnamefont {M.}~\bibnamefont
  {Cococcioni}}, \bibinfo {author} {\bibfnamefont {I.}~\bibnamefont {Dabo}},
  \bibinfo {author} {\bibfnamefont {A.}~\bibnamefont {{Dal Corso}}}, \bibinfo
  {author} {\bibfnamefont {S.}~\bibnamefont {de~Gironcoli}}, \bibinfo {author}
  {\bibfnamefont {S.}~\bibnamefont {Fabris}}, \bibinfo {author} {\bibfnamefont
  {G.}~\bibnamefont {Fratesi}}, \bibinfo {author} {\bibfnamefont
  {R.}~\bibnamefont {Gebauer}}, \bibinfo {author} {\bibfnamefont
  {U.}~\bibnamefont {Gerstmann}}, \bibinfo {author} {\bibfnamefont
  {C.}~\bibnamefont {Gougoussis}}, \bibinfo {author} {\bibfnamefont
  {A.}~\bibnamefont {Kokalj}}, \bibinfo {author} {\bibfnamefont
  {M.}~\bibnamefont {Lazzeri}}, \bibinfo {author} {\bibfnamefont
  {L.}~\bibnamefont {Martin-Samos}}, \bibinfo {author} {\bibfnamefont
  {N.}~\bibnamefont {Marzari}}, \bibinfo {author} {\bibfnamefont
  {F.}~\bibnamefont {Mauri}}, \bibinfo {author} {\bibfnamefont
  {R.}~\bibnamefont {Mazzarello}}, \bibinfo {author} {\bibfnamefont
  {S.}~\bibnamefont {Paolini}}, \bibinfo {author} {\bibfnamefont
  {A.}~\bibnamefont {Pasquarello}}, \bibinfo {author} {\bibfnamefont
  {L.}~\bibnamefont {Paulatto}}, \bibinfo {author} {\bibfnamefont
  {C.}~\bibnamefont {Sbraccia}}, \bibinfo {author} {\bibfnamefont
  {S.}~\bibnamefont {Scandolo}}, \bibinfo {author} {\bibfnamefont
  {G.}~\bibnamefont {Sclauzero}}, \bibinfo {author} {\bibfnamefont {A.~P.}\
  \bibnamefont {Seitsonen}}, \bibinfo {author} {\bibfnamefont {A.}~\bibnamefont
  {Smogunov}}, \bibinfo {author} {\bibfnamefont {P.}~\bibnamefont {Umari}}, \
  and\ \bibinfo {author} {\bibfnamefont {R.~M.}\ \bibnamefont {Wentzcovitch}},\
  }\href
  {http://iopscience.iop.org/article/10.1088/0953-8984/21/39/395502/meta}
  {\bibfield  {journal} {\bibinfo  {journal} {J. Phys.: Condens. Matter}\
  }\textbf {\bibinfo {volume} {21}},\ \bibinfo {pages} {395502} (\bibinfo
  {year} {2009})}\BibitemShut {NoStop}%
\bibitem [{\citenamefont {S\'{a}nchez-Portal}\ \emph
  {et~al.}(1996)\citenamefont {S\'{a}nchez-Portal}, \citenamefont {Artacho},\
  and\ \citenamefont {Soler}}]{Sanchez-Portal1996}%
  \BibitemOpen
  \bibfield  {author} {\bibinfo {author} {\bibfnamefont {D.}~\bibnamefont
  {S\'{a}nchez-Portal}}, \bibinfo {author} {\bibfnamefont {E.}~\bibnamefont
  {Artacho}}, \ and\ \bibinfo {author} {\bibfnamefont {J.~M.}\ \bibnamefont
  {Soler}},\ }\href
  {http://iopscience.iop.org/article/10.1088/0953-8984/8/21/012} {\bibfield
  {journal} {\bibinfo  {journal} {J. Phys.: Condens. Matter}\ }\textbf
  {\bibinfo {volume} {8}},\ \bibinfo {pages} {3859} (\bibinfo {year}
  {1996})}\BibitemShut {NoStop}%
\bibitem [{\citenamefont {Perdew}\ and\ \citenamefont
  {Zunger}(1981)}]{Perdew1981LDA}%
  \BibitemOpen
  \bibfield  {author} {\bibinfo {author} {\bibfnamefont {J.~P.}\ \bibnamefont
  {Perdew}}\ and\ \bibinfo {author} {\bibfnamefont {A.}~\bibnamefont
  {Zunger}},\ }\href {\doibase 10.1103/PhysRevB.23.5048} {\bibfield  {journal}
  {\bibinfo  {journal} {Phys. Rev. B}\ }\textbf {\bibinfo {volume} {23}},\
  \bibinfo {pages} {5048} (\bibinfo {year} {1981})}\BibitemShut {NoStop}%
\bibitem [{\citenamefont {Perdew}\ \emph {et~al.}(1996)\citenamefont {Perdew},
  \citenamefont {Burke},\ and\ \citenamefont {Ernzerhof}}]{Perdew1996}%
  \BibitemOpen
  \bibfield  {author} {\bibinfo {author} {\bibfnamefont {J.~P.}\ \bibnamefont
  {Perdew}}, \bibinfo {author} {\bibfnamefont {K.}~\bibnamefont {Burke}}, \
  and\ \bibinfo {author} {\bibfnamefont {M.}~\bibnamefont {Ernzerhof}},\ }\href
  {\doibase 10.1103/PhysRevLett.77.3865} {\bibfield  {journal} {\bibinfo
  {journal} {Phys. Rev. Lett.}\ }\textbf {\bibinfo {volume} {77}},\ \bibinfo
  {pages} {3865} (\bibinfo {year} {1996})}\BibitemShut {NoStop}%
\bibitem [{\citenamefont {Troullier}\ and\ \citenamefont
  {Martins}(1991)}]{Troullier1991NCPP}%
  \BibitemOpen
  \bibfield  {author} {\bibinfo {author} {\bibfnamefont {N.}~\bibnamefont
  {Troullier}}\ and\ \bibinfo {author} {\bibfnamefont {J.~L.}\ \bibnamefont
  {Martins}},\ }\href {\doibase 10.1103/PhysRevB.43.1993} {\bibfield  {journal}
  {\bibinfo  {journal} {Phys. Rev. B}\ }\textbf {\bibinfo {volume} {43}},\
  \bibinfo {pages} {1993} (\bibinfo {year} {1991})}\BibitemShut {NoStop}%
\bibitem [{\citenamefont {Kleinman}\ and\ \citenamefont
  {Bylander}(1982)}]{Kleinman1982SeparableNCPP}%
  \BibitemOpen
  \bibfield  {author} {\bibinfo {author} {\bibfnamefont {L.}~\bibnamefont
  {Kleinman}}\ and\ \bibinfo {author} {\bibfnamefont {D.~M.}\ \bibnamefont
  {Bylander}},\ }\href {\doibase 10.1103/PhysRevLett.48.1425} {\bibfield
  {journal} {\bibinfo  {journal} {Phys. Rev. Lett.}\ }\textbf {\bibinfo
  {volume} {48}},\ \bibinfo {pages} {1425} (\bibinfo {year}
  {1982})}\BibitemShut {NoStop}%
\bibitem [{\citenamefont {Dal~Corso}(2014)}]{DalCorso2014Pseudos}%
  \BibitemOpen
  \bibfield  {author} {\bibinfo {author} {\bibfnamefont {A.}~\bibnamefont
  {Dal~Corso}},\ }\href {\doibase
  http://dx.doi.org/10.1016/j.commatsci.2014.07.043} {\bibfield  {journal}
  {\bibinfo  {journal} {Comput. Mater. Sci}\ }\textbf {\bibinfo {volume}
  {95}},\ \bibinfo {pages} {337 } (\bibinfo {year} {2014})}\BibitemShut
  {NoStop}%
\bibitem [{Note3()}]{Note3}%
  \BibitemOpen
  \bibinfo {note} {See Fig.~9d in Ref.~\protect \rev@citealpnum
  {Giustino2007EPW}, which, as later clarified in Ref.~\protect \rev@citealpnum
  {Ponce2016EPW}, corresponds to the result for diamond.}\BibitemShut {Stop}%
\bibitem [{\citenamefont {Valiev}\ \emph {et~al.}(2010)\citenamefont {Valiev},
  \citenamefont {Bylaska}, \citenamefont {Govind}, \citenamefont {Kowalski},
  \citenamefont {Straatsma}, \citenamefont {Van~Dam}, \citenamefont {Wang},
  \citenamefont {Nieplocha}, \citenamefont {Apra}, \citenamefont {Windus},\
  and\ \citenamefont {de~Jong}}]{nwchem}%
  \BibitemOpen
  \bibfield  {author} {\bibinfo {author} {\bibfnamefont {M.}~\bibnamefont
  {Valiev}}, \bibinfo {author} {\bibfnamefont {E.~J.}\ \bibnamefont {Bylaska}},
  \bibinfo {author} {\bibfnamefont {N.}~\bibnamefont {Govind}}, \bibinfo
  {author} {\bibfnamefont {K.}~\bibnamefont {Kowalski}}, \bibinfo {author}
  {\bibfnamefont {T.~P.}\ \bibnamefont {Straatsma}}, \bibinfo {author}
  {\bibfnamefont {H.~J.}\ \bibnamefont {Van~Dam}}, \bibinfo {author}
  {\bibfnamefont {D.}~\bibnamefont {Wang}}, \bibinfo {author} {\bibfnamefont
  {J.}~\bibnamefont {Nieplocha}}, \bibinfo {author} {\bibfnamefont
  {E.}~\bibnamefont {Apra}}, \bibinfo {author} {\bibfnamefont {T.~L.}\
  \bibnamefont {Windus}}, \ and\ \bibinfo {author} {\bibfnamefont {W.~A.}\
  \bibnamefont {de~Jong}},\ }\href {\doibase
  https://doi.org/10.1016/j.cpc.2010.04.018} {\bibfield  {journal} {\bibinfo
  {journal} {Comput. Physics Commun.}\ }\textbf {\bibinfo {volume} {181}},\
  \bibinfo {pages} {1477} (\bibinfo {year} {2010})}\BibitemShut {NoStop}%
\bibitem [{\citenamefont {Souza}\ \emph {et~al.}(2001)\citenamefont {Souza},
  \citenamefont {Marzari},\ and\ \citenamefont {Vanderbilt}}]{Souza2001}%
  \BibitemOpen
  \bibfield  {author} {\bibinfo {author} {\bibfnamefont {I.}~\bibnamefont
  {Souza}}, \bibinfo {author} {\bibfnamefont {N.}~\bibnamefont {Marzari}}, \
  and\ \bibinfo {author} {\bibfnamefont {D.}~\bibnamefont {Vanderbilt}},\
  }\href {\doibase 10.1103/PhysRevB.65.035109} {\bibfield  {journal} {\bibinfo
  {journal} {Phys. Rev. B}\ }\textbf {\bibinfo {volume} {65}},\ \bibinfo
  {pages} {035109} (\bibinfo {year} {2001})}\BibitemShut {NoStop}%
\bibitem [{\citenamefont {Maradudin}\ and\ \citenamefont
  {Vosko}(1968)}]{Maradudin1968}%
  \BibitemOpen
  \bibfield  {author} {\bibinfo {author} {\bibfnamefont {A.~A.}\ \bibnamefont
  {Maradudin}}\ and\ \bibinfo {author} {\bibfnamefont {S.~H.}\ \bibnamefont
  {Vosko}},\ }\href {\doibase 10.1103/RevModPhys.40.1} {\bibfield  {journal}
  {\bibinfo  {journal} {Rev. Mod. Phys.}\ }\textbf {\bibinfo {volume} {40}},\
  \bibinfo {pages} {1} (\bibinfo {year} {1968})}\BibitemShut {NoStop}%
\bibitem [{\citenamefont {Ponc{\'e}}\ \emph {et~al.}(2016)\citenamefont
  {Ponc{\'e}}, \citenamefont {Margine}, \citenamefont {Verdi},\ and\
  \citenamefont {Giustino}}]{Ponce2016EPW}%
  \BibitemOpen
  \bibfield  {author} {\bibinfo {author} {\bibfnamefont {S.}~\bibnamefont
  {Ponc{\'e}}}, \bibinfo {author} {\bibfnamefont {E.}~\bibnamefont {Margine}},
  \bibinfo {author} {\bibfnamefont {C.}~\bibnamefont {Verdi}}, \ and\ \bibinfo
  {author} {\bibfnamefont {F.}~\bibnamefont {Giustino}},\ }\href {\doibase
  https://doi.org/10.1016/j.cpc.2016.07.028} {\bibfield  {journal} {\bibinfo
  {journal} {Comput. Phys. Commun.}\ }\textbf {\bibinfo {volume} {209}},\
  \bibinfo {pages} {116 } (\bibinfo {year} {2016})}\BibitemShut {NoStop}%
\end{thebibliography}

\begin{thebibliography}{4}%
\makeatletter
\providecommand \@ifxundefined [1]{%
 \@ifx{#1\undefined}
}%
\providecommand \@ifnum [1]{%
 \ifnum #1\expandafter \@firstoftwo
 \else \expandafter \@secondoftwo
 \fi
}%
\providecommand \@ifx [1]{%
 \ifx #1\expandafter \@firstoftwo
 \else \expandafter \@secondoftwo
 \fi
}%
\providecommand \natexlab [1]{#1}%
\providecommand \enquote  [1]{``#1''}%
\providecommand \bibnamefont  [1]{#1}%
\providecommand \bibfnamefont [1]{#1}%
\providecommand \citenamefont [1]{#1}%
\providecommand \href@noop [0]{\@secondoftwo}%
\providecommand \href [0]{\begingroup \@sanitize@url \@href}%
\providecommand \@href[1]{\@@startlink{#1}\@@href}%
\providecommand \@@href[1]{\endgroup#1\@@endlink}%
\providecommand \@sanitize@url [0]{\catcode `\\12\catcode `\$12\catcode
  `\&12\catcode `\#12\catcode `\^12\catcode `\_12\catcode `\%12\relax}%
\providecommand \@@startlink[1]{}%
\providecommand \@@endlink[0]{}%
\providecommand \url  [0]{\begingroup\@sanitize@url \@url }%
\providecommand \@url [1]{\endgroup\@href {#1}{\urlprefix }}%
\providecommand \urlprefix  [0]{URL }%
\providecommand \Eprint [0]{\href }%
\providecommand \doibase [0]{http://dx.doi.org/}%
\providecommand \selectlanguage [0]{\@gobble}%
\providecommand \bibinfo  [0]{\@secondoftwo}%
\providecommand \bibfield  [0]{\@secondoftwo}%
\providecommand \translation [1]{[#1]}%
\providecommand \BibitemOpen [0]{}%
\providecommand \bibitemStop [0]{}%
\providecommand \bibitemNoStop [0]{.\EOS\space}%
\providecommand \EOS [0]{\spacefactor3000\relax}%
\providecommand \BibitemShut  [1]{\csname bibitem#1\endcsname}%
\let\auto@bib@innerbib\@empty
\bibitem [{\citenamefont {Marzari}\ \emph {et~al.}(1994)\citenamefont
  {Marzari}, \citenamefont {de~Gironcoli},\ and\ \citenamefont
  {Baroni}}]{Marzari1994VirtualCrystal}%
  \BibitemOpen
  \bibfield  {author} {\bibinfo {author} {\bibfnamefont {N.}~\bibnamefont
  {Marzari}}, \bibinfo {author} {\bibfnamefont {S.}~\bibnamefont
  {de~Gironcoli}}, \ and\ \bibinfo {author} {\bibfnamefont {S.}~\bibnamefont
  {Baroni}},\ }\href {\doibase 10.1103/PhysRevLett.72.4001} {\bibfield
  {journal} {\bibinfo  {journal} {Phys. Rev. Lett.}\ }\textbf {\bibinfo
  {volume} {72}},\ \bibinfo {pages} {4001} (\bibinfo {year}
  {1994})}\BibitemShut {NoStop}%
\bibitem [{\citenamefont {Boeri}\ \emph {et~al.}(2004)\citenamefont {Boeri},
  \citenamefont {Kortus},\ and\ \citenamefont
  {Andersen}}]{Boeri2004BoronDiamond}%
  \BibitemOpen
  \bibfield  {author} {\bibinfo {author} {\bibfnamefont {L.}~\bibnamefont
  {Boeri}}, \bibinfo {author} {\bibfnamefont {J.}~\bibnamefont {Kortus}}, \
  and\ \bibinfo {author} {\bibfnamefont {O.~K.}\ \bibnamefont {Andersen}},\
  }\href {\doibase 10.1103/PhysRevLett.93.237002} {\bibfield  {journal}
  {\bibinfo  {journal} {Phys. Rev. Lett.}\ }\textbf {\bibinfo {volume} {93}},\
  \bibinfo {pages} {237002} (\bibinfo {year} {2004})}\BibitemShut {NoStop}%
\bibitem [{\citenamefont {Giustino}\ \emph
  {et~al.}(2007{\natexlab{a}})\citenamefont {Giustino}, \citenamefont {Yates},
  \citenamefont {Souza}, \citenamefont {Cohen},\ and\ \citenamefont
  {Louie}}]{Giustino2007BoronDiamond}%
  \BibitemOpen
  \bibfield  {author} {\bibinfo {author} {\bibfnamefont {F.}~\bibnamefont
  {Giustino}}, \bibinfo {author} {\bibfnamefont {J.~R.}\ \bibnamefont {Yates}},
  \bibinfo {author} {\bibfnamefont {I.}~\bibnamefont {Souza}}, \bibinfo
  {author} {\bibfnamefont {M.~L.}\ \bibnamefont {Cohen}}, \ and\ \bibinfo
  {author} {\bibfnamefont {S.~G.}\ \bibnamefont {Louie}},\ }\href {\doibase
  10.1103/PhysRevLett.98.047005} {\bibfield  {journal} {\bibinfo  {journal}
  {Phys. Rev. Lett.}\ }\textbf {\bibinfo {volume} {98}},\ \bibinfo {pages}
  {047005} (\bibinfo {year} {2007}{\natexlab{a}})}\BibitemShut {NoStop}%
\bibitem [{\citenamefont {Giustino}\ \emph
  {et~al.}(2007{\natexlab{b}})\citenamefont {Giustino}, \citenamefont {Cohen},\
  and\ \citenamefont {Louie}}]{Giustino2007EPW}%
  \BibitemOpen
  \bibfield  {author} {\bibinfo {author} {\bibfnamefont {F.}~\bibnamefont
  {Giustino}}, \bibinfo {author} {\bibfnamefont {M.~L.}\ \bibnamefont {Cohen}},
  \ and\ \bibinfo {author} {\bibfnamefont {S.~G.}\ \bibnamefont {Louie}},\
  }\href {\doibase 10.1103/PhysRevB.76.165108} {\bibfield  {journal} {\bibinfo
  {journal} {Phys. Rev. B}\ }\textbf {\bibinfo {volume} {76}},\ \bibinfo
  {pages} {165108} (\bibinfo {year} {2007}{\natexlab{b}})}\BibitemShut
  {NoStop}%
\end{thebibliography}
%

%
%
\widetext
\newpage
\begin{center}
\textbf{\Large Supplemental Material:\\ \textit{Ab Initio} Electron-Phonon Interactions Using Atomic Orbital Wavefunctions}\\
\vspace{5mm}
Luis A. Agapito and Marco Bernardi\\
Department of Applied Physics and Materials Science, Steele Laboratory, California Institute of Technology, Pasadena, California 91125, United States
\end{center}
\setcounter{equation}{0}
\setcounter{figure}{0}
\setcounter{table}{0}
\setcounter{page}{1}
\setcounter{section}{0}
\makeatletter
\renewcommand{\theequation}{S\arabic{equation}}
\renewcommand{\thefigure}{S\arabic{figure}}
\renewcommand{\thetable}{S\arabic{table}}

\section*{1. Band structure interpolation in silicon using\\ atomic orbitals and Wannier functions}
\begin{figure}[h]
\centering
\includegraphics[width=8.6cm]{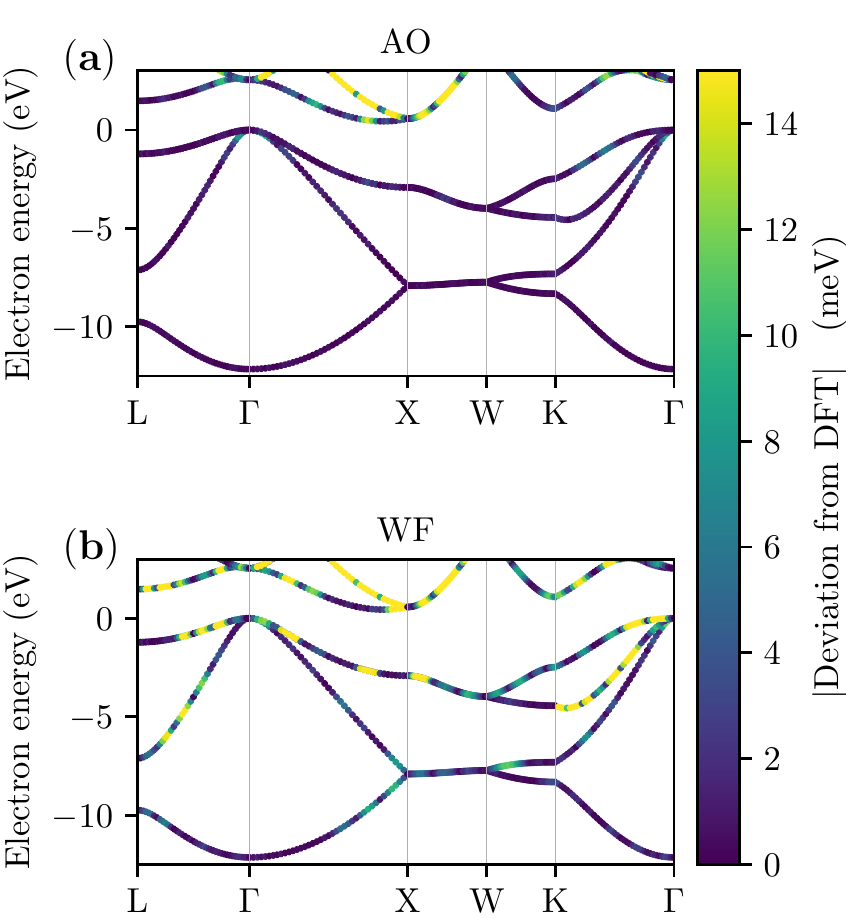}
\caption{Deviations from the DFT eigenvalues of the electronic eigenvalues obtained using AO (a) and WF (b) interpolations. Both interpolation methods employ Bloch electronic wavefunctions evaluated at a coarse $12 \times12 \times12$ $\bk$-point grid. The accuracy of the two methods is comparable.} 
\label{fig:fig_bandscompare}
\end{figure}

\section*{2. Boron-doped diamond}
\begin{figure}[h]
\centering
\includegraphics[width=14cm]{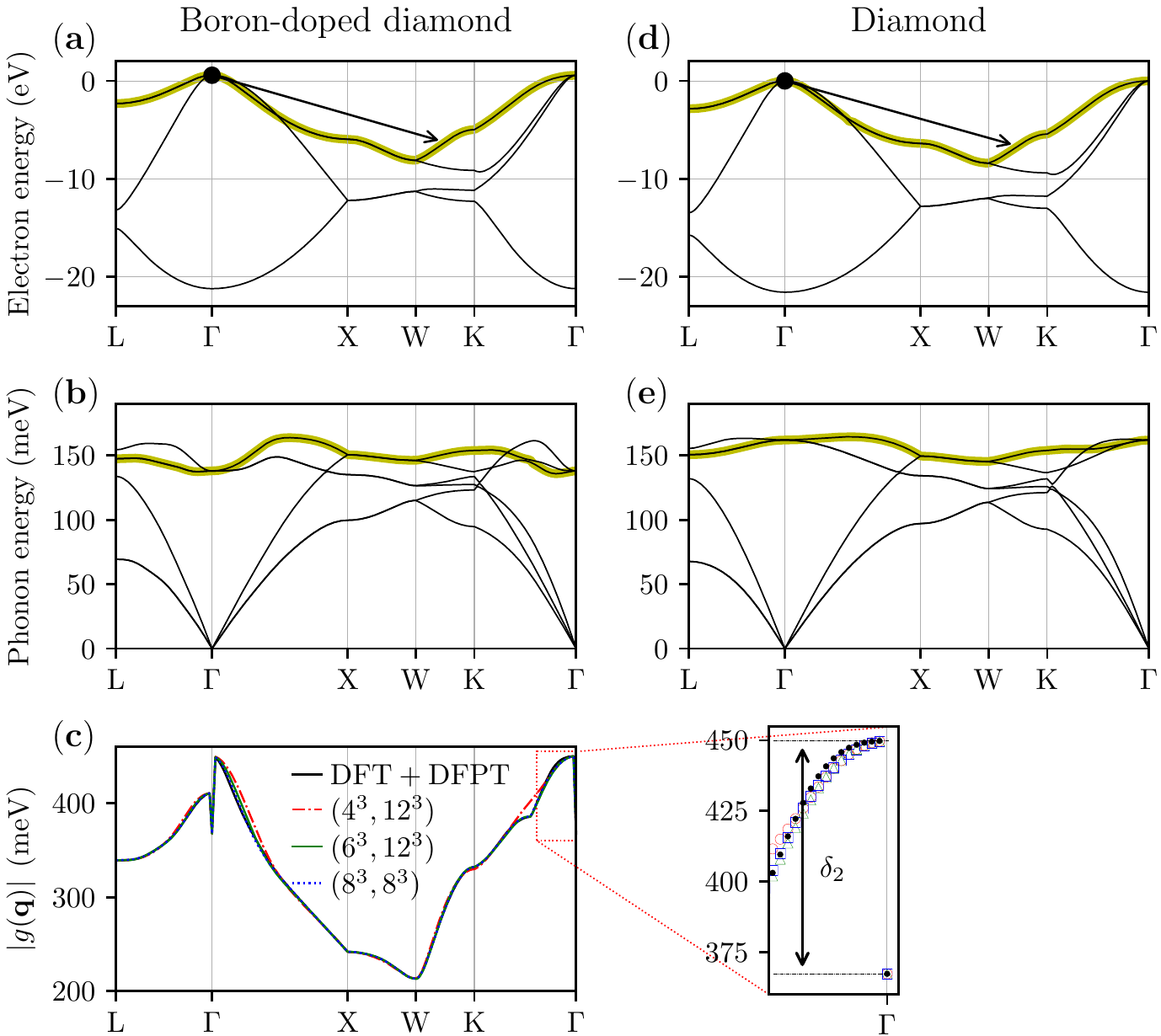}
\caption{Left panels: Interpolation of the e-ph matrix elements for boron-doped diamond using WFs. The selected electronic band and phonon mode are highlighted in (a) and (b), respectively. 
The DFT plus DFPT and interpolated e-ph matrix elements, for several ($\bq_{\rm c}$, $\bk_{\rm c}$) coarse grids, are given in (c). The inset zooms into the region near $\Gamma$. Right panels: the electronic band structure and phonon 
dispersions of pristine diamond are given for comparison.} 
\label{fig:fig_doped}
\end{figure}

We use a diamond unit cell together with the virtual crystal approximation \citesinfo{Marzari1994VirtualCrystal} to model the doping of diamond with boron. We employ a virtual atom $\expval{\textrm{BC}}$ at every site with a composite pseudopotential that is the weighted average between those of boron and carbon. The norm-conserving pseudopotential of the virtual atoms is 
\begin{equation}
	\hat{V}^{\expval{\textrm{BC}}} = x \hat{V}_{\rm loc}^{\textrm{B}} + (1-x) \hat{V}_{\rm loc}^{\textrm{C}} + \sum_{ij} \ket{\beta^{\textrm{B}}_i} x D^{\textrm{B}}_{ij} \bra{\beta^{\textrm{B}}_j} + \sum_{ij} \ket{\beta^{\textrm{C}}_i}(1-x)D^{\textrm{C}}_{ij} \bra{\beta^{\textrm{C}}_j},
\end{equation}
where $\hat{V}_{\rm loc}$ is the local part of the pseudopotential, the nonlocal part is given in the Kleinman-Bylander form through the beta projectors $\ket{\beta_i}$ and coupling coefficients $D_{ij}$, and $x$ is the boron fraction. 
We use a doping of one boron atom every 54 carbon atoms, which corresponds to $x=0.0185185$. 
The DFT calculation shows a rigid shift of the band structure [Fig.~\ref{fig:fig_doped}(a)] with respect to the undoped case [Fig.~\ref{fig:fig_doped}(d)]; the system becomes slightly metallic, with the Fermi level located 0.58 eV below the top of the valence band. The phonon dispersions show that the doping affects mostly the acoustic modes around $\Gamma$ [Figs.~\ref{fig:fig_doped}(b) and (e)], in agreement with previous reports~\citesinfo{Boeri2004BoronDiamond,Giustino2007BoronDiamond,Giustino2007EPW}.

Fig.~\ref{fig:fig_doped}(c) shows the interpolated e-ph matrix elements for boron-doped diamond, computed using different coarse $\bq$-point grids and using the electronic states and phonon modes highlighted in Figs.~\ref{fig:fig_doped}(a) and (b). Different from undoped diamond, which is shown in Fig.~\ref{fig:fig_gconverge}(d) of the main text, we find a rapid convergence with increasing coarse $\bq$-point grid density of the interpolated e-ph matrix elements near $\Gamma$ along the K$-$$\Gamma$ direction [see the inset in Fig.~\ref{fig:fig_doped}(c)]. This behavior is due to the metallic character of boron-doped diamond, which eliminates the DFPT discontinuity at $\bq\!=\!0$, as explained in the main text. Note that the discontinuity $\delta_2$ near $\Gamma$ is still present, since it is due to electron and phonon degeneracies, as discussed in the main text. If the same phonon modes as those in Fig. 8 of Ref.~\onlinecite{Giustino2007EPW} are chosen (not shown), 
we obtain interpolated e-ph matrix elements in excellent agreement with Ref.~\onlinecite{Giustino2007EPW}. 

\begin{table}[h]
\caption{Difference (in meV units) between the DFT plus DFPT and the interpolated e-ph matrix elements in diamond at three high-symmetry points. 
The AO and WF interpolated results are given for several coarse grids.}
\label{table:C-AOvsWF} 
\begin{tabular*}{0.5\textwidth}{c|c|c|c|c}
\hline
\hline
$\bqf$ point & Method &\multicolumn{3}{c}{Coarse grid size ($\bqc$ grid, $\bkc$ grid)}\\
        &        & ($4^3$, $12^3$) & ($6^3$, $12^3$) & ($8^3$, $8^3$)\\
\hline
\multirow{2}{*}{K = [-$\frac{3}{8}$, $\frac{3}{8}$, 0]}&AO&  2.51 &  3.39 &  3.30\\ 
                                                       &WF& -0.59 &  0.17 &  0.00\\
\hline
\multirow{2}{*}{\!\!\!\!\!L = [0, $\frac{1}{2}$,0]}              &AO& -5.31 & -5.31 & -5.31\\ 
                                                       &WF&  0.00 &  0.00 &  0.00  \\
\hline
\multirow{2}{*}{X = [0, $\frac{1}{2}$, $\frac{1}{2}$]} &AO& -7.06 & -7.06 & -7.06\\ 
                                                       &WF&  0.00 &  0.00 &  0.00\\
\hline
\hline
\end{tabular*}
\end{table}
%
%

\end{document}